\documentclass[fleqn,usenatbib]{mnras}

\usepackage{newtxtext,newtxmath}

\usepackage[T1]{fontenc}

\DeclareRobustCommand{\VAN}[3]{#2}
\let\VANthebibliography\thebibliography
\def\thebibliography{\DeclareRobustCommand{\VAN}[3]{##3}\VANthebibliography}

\usepackage{graphicx}	
\usepackage{amsmath}	
\usepackage{xcolor}
\definecolor{MyBlue}{RGB}{0,0,255}
\usepackage{float}
\usepackage{hyperref}
\usepackage{subfigure}
\usepackage{orcidlink}
\newcommand{\suporcid}[1]{\textsuperscript{\orcidlink{#1}}}

\title[OJ 287: Multi-Epoch SED Study]{Transitional Spectral Behavior in Blazar OJ 287: A Comprehensive Multi-Epoch SED Study}

\author[Navaneeth et al.]{P. K. Navaneeth\suporcid{0000-0001-9092-2903},$^{1}$ Gopal Bhatta\suporcid{0000-0002-0705-6619},$^{1}$ Sangeetha Kizhakkekalam\suporcid{0009-0000-4001-2012}$^{1}$
\\
$^{1}$Janusz Gil Institute of Astronomy, University of Zielona Góra, ul. Szafrana 2, 65-516 Zielona Góra, Poland}

\date{Accepted XXX. Received YYY; in original form ZZZ}

\pubyear{\the\year{}}

\begin{document}
\label{firstpage}
\pagerange{\pageref{firstpage}--\pageref{lastpage}}
\maketitle

\begin{abstract}
We present a comprehensive multi-wavelength investigation of BL Lac object OJ 287 using Swift and Fermi observatories spanning 2008-2025. The source exhibits significant flux variability across optical-UV, X-ray, and $\gamma$-ray regimes, with outbursts observed in optical, UV and X-ray frequencies. Variability and correlation analyses of the long-term lightcurves reveal strong correlations among optical, UV, and X-ray bands, whereas there is no significant correlation between $\gamma$-ray and other bands. Using a decade-long dataset (MJD 57382-60448), we analysed X-ray spectral variability across different flux states (flare, intermediate, and quiescent). Multi-wavelength spectral energy distributions (SEDs) were constructed using a one-zone leptonic model to investigate the transitional nature of the X-ray spectral position in the broadband. The analysis reveals OJ 287's consistent flux-dependent transitional spectral behavior: quiescent states show hard X-ray spectra dominated by inverse Compton emission; intermediate states exhibit contributions from both inverse Compton and synchrotron components with moderately hard spectra; flaring events are characterized by predominantly synchrotron emission, resulting in soft X-ray spectra. Broadband SED modelling captures the systematic evolution of model parameters across different activity states, with correlation analysis revealing strong positive correlations between synchrotron peak frequency, Synchrotron-Self Compton peak frequency, and X-ray flux, providing quantitative evidence for the changing dominance of emission components within the blazar's relativistic jet.
\end{abstract}

\begin{keywords}
galaxies: active -  galaxies: BL Lacertae objects: individual: OJ 287 - X-rays: galaxies - gamma-rays: galaxies - galaxies: jets - radiation mechanisms: non-thermal
\end{keywords}


\section{Introduction} \label{sec:intro}
Active galactic nuclei (AGN) are the most powerful accretion-powered sources in the universe. Blazars, a subclass of AGN, are distinguished by their highly relativistic jets oriented within a narrow angle of $\leq 10^\circ$ toward our line of sight \citep{1995PASP..107..803U}. These sources exhibit high variability in luminosity, flux, and polarization across timescales, distinguishing them from other astrophysical objects. Blazars show multi-wavelength (MWL) emission with significantly Doppler-boosted radiation spanning the entire spectrum from radio wavelengths to very high energy (VHE) $\gamma$-rays \citep[see e.g.,][]{2019Galax...7...20B,2022Univ....8..513B}. The blazar population is conventionally divided into two primary subclasses based on their optical spectral characteristics: Flat Spectrum Radio Quasars (FSRQs) and BL Lacertae objects (BL Lacs). FSRQs are distinguished by the presence of prominent broad emission lines in their optical spectra \citep{blandford1978extended, ghisellini1997optical}, while BL Lacs exhibit either absent or weak emission line features \citep{stocke1991einstein, marcha1996optical}. BL Lac objects are less powerful, often found to be TeV emitters, with their inverse Compton peak occurring at the highest $\gamma$-ray energies. FSRQs, although powerful blazars, are dominated by Compton emission, with the synchrotron peak occurring at lower frequencies \citep{bhatta2024gamma}.  Blazars are also classified based on the location of their synchrotron peak: High Synchrotron Peaked blazars (HSP; $\nu_s > 10^{15}$ Hz), Intermediate Synchrotron Peaked blazars (ISP; $10^{14} < \nu_s < 10^{15}$ Hz), and Low Synchrotron Peaked blazars (LSP; $\nu_s < 10^{14}$ Hz) \citep{abdo2010fermi}.

The MWL SED of blazars characteristically exhibits a distinctive double-peaked structure. The lower-energy component reaches its maximum in the optical to soft X-ray regime and originates from synchrotron radiation produced by relativistic particles within the jet's magnetic field environment. The physical mechanism responsible for the higher-energy peak, which extends into the MeV-TeV range, remains a subject of ongoing theoretical investigation \citep{2019ApJ...887..133B}. 

Theoretical frameworks propose two principal emission mechanisms to account for blazar radiation. In leptonic models, the dominant radiative processes are attributed to relativistic leptons, such as electrons and positrons, by inverse-Compton scattering of low-energy seed photons. The source of these seed photons varies: They originate from synchrotron radiation within the jet itself in the Synchrotron self-Compton (SSC) scenario \citep{maraschi1992jet}, whereas in the External-Compton (EC) model, they originate from the accretion disk \citep{dermer1993model}, the broad line region \citep{sikora1994comptonization}, or the dusty torus \citep{blazejowski2000comptonization}.  The hadronic models propose that the high-energy emission results from synchrotron radiation by ultra-relativistic protons or secondary leptons generated through proton-proton interactions \citep[see e.g.,][]{2001APh....15..121M, 2013ApJ...768...54B}.

In this work, we  present a comprehensive MWL timing and spectral analysis of the blazar OJ 287, utilizing data from the Neil Gehrels Swift Observatory and the Fermi Large Area Telescope (LAT). The structure of the paper is organized as follows: Section \ref{sec:source} introduces OJ 287, including a review of significant previous investigations of this source; Section \ref{sec:data} describes the MWL data reduction and analysis procedures; Section \ref{sec:analysis} outlines the analytical methodology and presents the results; Section \ref{sec:discussion} discusses the findings and their implications; and finally, Section \ref{sec:summary} summarizes the main conclusions of this study.

\section{Blazar OJ 287} \label{sec:source}
OJ 287 is a bright and highly variable BL Lac object located at z = 0.306, discovered in 1967 \citep{dickel1967survey}. The source is one of the best candidates for the characterization of the BL Lacertae class of sources, mostly due to its highly dynamic and MWL variability, along with observational properties, such as optical and radio brightness \citep{sitko1985continuum}. Over the past five decades, OJ 287 has demonstrated remarkable variability behavior \citep{andrew1971oj}. A systematic analysis of its variability across different time scales was conducted by \citep{valtonen20062005}, as discussed by \citep{valtaoja198515}, which provides a detailed study of intraday variability in both radio and optical observations. The source has also been extensively studied in X-ray \citep{2021MNRAS.508..315P,2022MNRAS.510.5280M,2024MNRAS.532.3285Z} and gamma-ray \citep{2020MNRAS.499..653K,2011MNRAS.412.1389N} observations, further showing its MWL variability characteristics.

The blazar OJ 287 is particularly renowned for exhibiting recurring patterns in its optical lightcurves spanning several decades, with a periodicity of approximately 12 years. This consistent recurrence prompted researchers to model the source as a binary black hole system, wherein a smaller black hole orbits a substantially larger primary, perturbing the latter's accretion disk \citep{sillanpaa1988oj}. The distinctive double-peaked flaring feature led to propose a scenario in which the secondary black hole periodically impacts the primary's accretion disk. This model, supported by decades of observational data and refined through seminal works such as \citet{1996ApJ...460..207L} and \citet{valtonen2016primary}, offers an explanation for the blazar's predictable optical outbursts. These outbursts have proven crucial for testing principles of general relativity, as highlighted by \citet{2008Natur.452..851V}, and are theorized to result from the secondary black hole's disk impacts. The model's predictive capability was compellingly showed by \citet{2018ApJ...866...11D} through the accurate forecast of the 2015 flare, substantially reinforcing its validity. Additional supporting evidence includes the precessing jet observed in high-resolution radio imaging \citep[e.g.,][]{2018MNRAS.478.3199B, 2022ApJ...924..122G}, further suggesting a binary interpretation. Nevertheless, the non-detection of the predicted 2022 outburst presents a significant challenge to this model. This discrepancy suggests that alternative binary supermassive black hole scenarios featuring a less massive primary black hole and reduced orbital precession may offer more plausible explanations \citep{2023MNRAS.522L..84K}. A beaming model for periodicity proposes that the interaction of two jets with the ambient medium causes them to bend, and when combined with long-term jet precession, this results in a periodically varying optical flux \citep{1998MNRAS.293L..13V}. In addition to the well-established 12-year periodicity, numerous studies have reported the detection of quasi-periodic oscillations (QPOs) in OJ 287 across various electromagnetic bands. These QPOs manifest on diverse timescales ranging from tens of days to months to years, depending on the temporal span of the analyzed data \citep{Pihajoki_2013, 2016ApJ...832...47B, 2020MNRAS.499..653K, Britzen_2023}.

High-resolution VLBA radio observations suggest that OJ 287 contains a parsec-scale jet, which appears to exhibit wobbling motion over time. The projected position angle of this jet varies by more than 100$^\circ$ \citep{tateyama2004structure, agudo2010location, moor2011connection}. Observations in the optical band indicate a preferred position angle, suggesting the presence of both a stable core and a turbulent jet component \citep{villforth2010variability}. Correlation analyses between gamma-ray and radio emission reveal connection with flaring events and the emergence of superluminal components from stationary knots along the jet, as seen in VLBA images \citep{agudo2010location, sawada2015apparent, hodgson2017location}. The source has been extensively studied through the MWL Observations and Modelling of OJ 287 (MOMO; \citealt{2021Univ....7..261K}) program, aimed to conduct comprehensive, long-term, and dense monitoring across multiple frequencies, ranging from radio to X-rays.

\section{Data Analysis} \label{sec:data}

\begin{figure*}
\includegraphics[width=\linewidth]{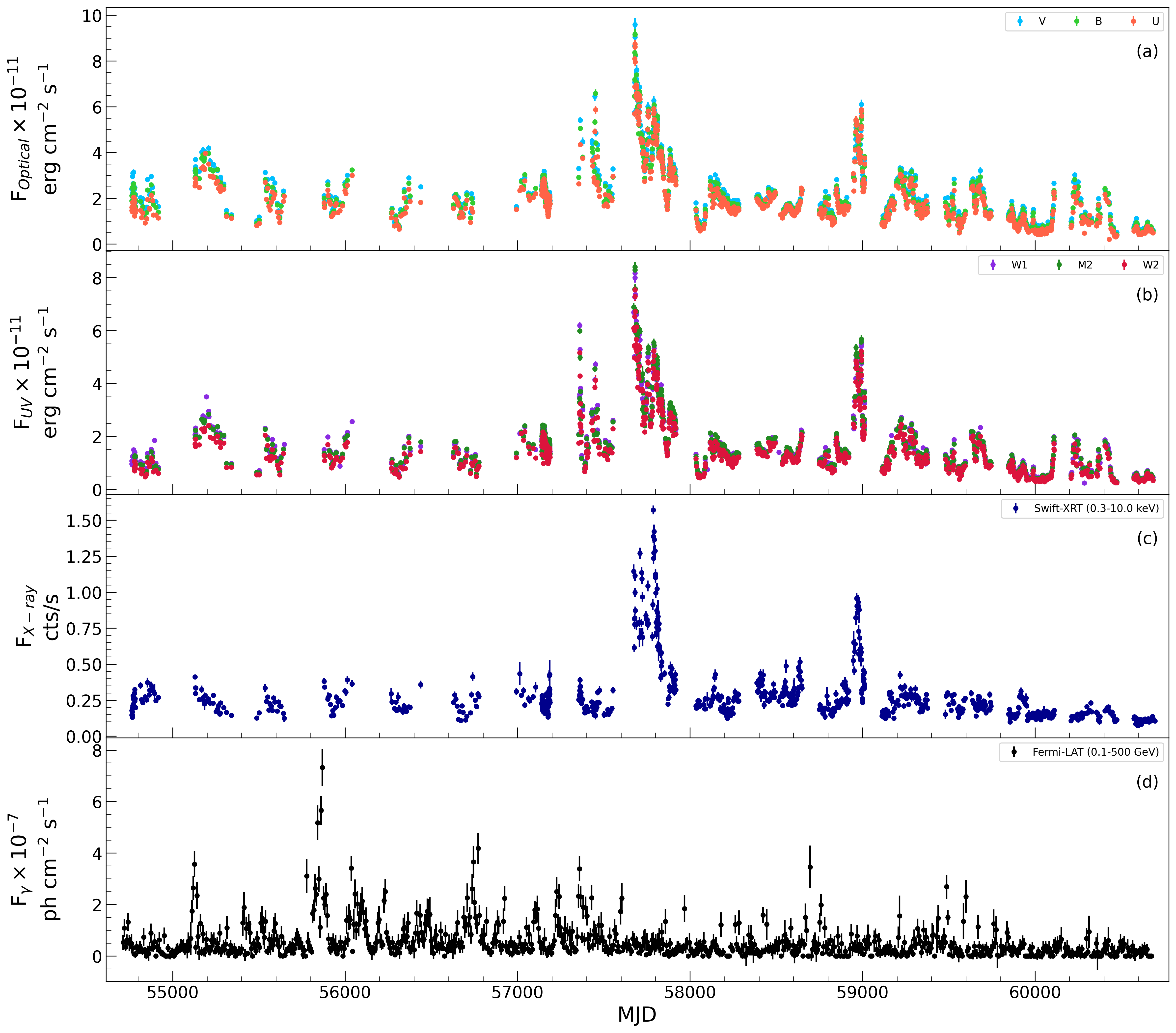}
\caption{Multi-wavelength lightcurves of the BL Lac OJ 287 spanning from MJD 54613 to 60775 are presented in panels from top to bottom: (a) Swift-UVOT Optical bands, (b) Swift-UVOT UV bands, (c) Swift-XRT X-ray, (d) Fermi-LAT $\gamma$-ray.
\label{fig:mwllc}}
\end{figure*}

\subsection{Fermi-LAT} 
The Large Area Telescope (LAT) aboard the Fermi Gamma-ray Space Telescope, launched by NASA in 2008 \citep{2009ApJ...697.1071A}, is a pair-conversion telescope that operates in scanning mode, surveying the entire sky approximately every 3 hours across the energy range of $\sim$20 MeV - 1 TeV. The data\footnote{\url{https://fermi.gsfc.nasa.gov/ssc/data/access/lat/}} were downloaded for a 20$^\circ$ region of interest (ROI) centered on the source position (RA: 133.7036, Dec: 20.1085). The analysis utilized Pass 8 data to search for gamma-ray emission in the energy range of 0.1–300 GeV, employing FermiTools 2.2.0 and Fermipy 1.2.2 \citep{2017ICRC...35..824W}.  The SOURCE class events (evclass = 128 and evtype = 3) were selected within a $10^{\circ}$ region of interest centered on the source position. Standard data quality cuts were applied, including a zenith angle cut of zmax $<$ 90° and the selection of good time intervals based on "\texttt{(DATA\_QUAL$>$0)\&\&(LAT\_CONFIG==1)}". Due to the rapidly decreasing effective area and increasing point-spread function (PSF) at low energies (E $<$ 100 MeV), all point sources from the 4FGL catalog within $15^{\circ}$ of the region of interest center were included in the initial model file, along with the Galactic (gll\_iem\_v07) and extragalactic isotropic diffuse emission (iso\_P8R3\_SOURCE\_V3\_v1) components. A binned likelihood analysis was performed, using eight energy bins per decade and a spatial binning of $0.1^{\circ}$ per pixel, to optimize the spectral parameters.  After initial optimization, we fixed sources with test statistics (TS) $<$ 10 and allowed the normalization of sources with TS $>$ 10 to vary in the model ﬁle. We additionally freed the normalization parameters for all sources within 3 degrees of the ROI center. Furthermore, all parameters associated with the isotropic and Galactic diffuse components were allowed to vary freely.

\subsection{Swift-XRT}
The XRT (X-ray Telescope) is part of the Swift Observatory \citep{2004ApJ...611.1005G} mission, operational since 2004. It is a focusing X-ray telescope with a 110 $cm^2$ effective area, 23$^{\prime}$ FOV, 18$^{\prime\prime}$ resolution (half-power diameter), and 0.2-10 keV energy range \citep{2005SSRv..120..165B}. While primarily designed for studying Gamma-Ray Bursts (GRBs) and their afterglows, XRT, which measures fluxes, spectra, and lightcurves over a wide dynamic range spanning more than 7 orders of magnitude in flux, has proven valuable for MWL studies of various compact objects and facilitates the detection and analysis of diverse X-ray sources.

The X-ray spectra and lightcurves of OJ 287 were generated using the online Swift-XRT data products generator tool\footnote{\url{https://www.swift.ac.uk/user_objects/}}. This automated online tool enables building XRT lightcurve, spectrum, image, enhanced position or source detection of any point source observed by Swift. The details of the data reduction process followed by this online tool can be found in \cite{2007A&A...469..379E, 2009MNRAS.397.1177E}. The 0.3-10 keV X-ray spectra of OJ 287 were rebinned to 20 minimum counts in each energy bin and fitted with an absorbed power-law (tbabs*powerlaw) model in XSPEC\footnote{\url{https://heasarc.gsfc.nasa.gov/xanadu/xspec/}} \citep{1996ASPC..101...17A} version 12.14.0. While fitting, we fixed the Galactic hydrogen column density (n$\rm _H$) to $1.28\times10^{20}$ $cm^{-2}$ \citep{2013MNRAS.431..394W}. 

\subsection{Swift-UVOT}
The UV/Optical Telescope (UVOT) is one of three instruments aboard the Swift Observatory \citep{2005SSRv..120...95R}. The UVOT instrument observed OJ 287 in its three optical (u, b and v) and three ultraviolet (w1, m2 and w2) filters. All the available UVOT observations for OJ 287 have been downloaded from the HEASARC archive\footnote{\url{https://heasarc.gsfc.nasa.gov/cgi-bin/W3Browse/w3browse.pl}} and performed data analysis using the HEASoft package version 6.33 and the CALDB version 20220331. We followed the standard procedures of data reduction given by the tutorials in The Swift UVOT Software Guide \citep{2008UVOTguide}. The source counts were extracted using a 6 arcsec radius centered on the target, while background counts were measured from a 20 arcsec radius circle placed in a nearby source-free region. Images across all extensions were combined for each observation using $uvotimsum$, performed separately for each filter. Then $uvot2pha$ tool was employed to generate spectral products for individual filters per observation. In our spectral analysis, we included only those UVOT observations where imaging data were available in a minimum of 3 filters. The selected spectra are then fitted using XSPEC with an absorbed power-law (zphabs*redden*powerlaw) model. The integrated fluxes were corrected for Galactic absorption by fixing the value of parameter E(B-V) magnitude to 0.018 \citep{2011ApJ...737..103S}. 

\section{Analysis and results} \label{sec:analysis}

\subsection{Multi-wavelength variability} \label{sec:mwlvar}
For timing analysis, we analysed $\sim$16 years of MWL data of OJ 287 from 2008 September to 2025 January in Optical-UV, X-rays and $\gamma$-rays. MWL lightcurves are plotted in Figure \ref{fig:mwllc}. The top panel displays the Swift-UVOT optical band lightcurves, while the second panel shows the Swift-UVOT ultraviolet bands. Both optical and ultraviolet data are binned by observation. The third panel presents the Swift-XRT X-ray lightcurve, also binned by observation. The bottom panel depicts the Fermi-LAT $\gamma$-ray lightcurve with weekly binning. The figure shows that OJ 287 exhibits strong flux variability in all four wavelength regimes. The MWL lightcurve analysis reveals strong correlation patterns among optical, ultraviolet, and X-ray bands. Two prominent X-ray flaring events were detected in 2017 and 2020, both exhibiting clear optical/UV counterparts but lacking corresponding $\gamma$-ray activity. Conversely, an optical/UV flare observed in 2015 without detectable signatures in either X-ray or $\gamma$-ray bands. The X-ray and optical/UV flares have been studied earlier by \cite{2020MNRAS.498L..35K}, \cite{2018MNRAS.479.1672K}, \cite{2018MNRAS.473.1145K} and \cite{2021A&A...654A..38P}. Detailed correlation analyses between X-ray emission and other wavelength regimes are presented in Section \ref{sec:corrzdcf}. The MWL investigation shows that the $\gamma$-ray emission exhibits no statistically significant correlation with other wavebands throughout the monitoring period. The historical $\gamma$-ray data reveal multiple distinct flaring events, with the most prominent outburst occurring between MJD 55809.0 and 55900.0, spanning 91 days. The flare does not have any counterparts in other wavebands. The highest flux observed in $\gamma$-rays is $7.32\times10^{-07}\pm 7.22\times10^{-08}$ $\text{ph cm}^{-2}\:\text{s}^{-1}$ at MJD 55868.5, coinciding with the flare maximum.

\subsection{Fractional Variability} \label{sec:fv}
The historical lightcurves distinctly show the variable nature of OJ 287. The average variability during the entire period can be quantified by estimating their fractional variability and is computed as:
\begin{equation} 
    F_{\rm var} = \sqrt{\frac{S^{2} - \bar{\sigma}^{2}_{\rm err}}{\bar{X}^2}},
    \label{eq:fv}
\end{equation}
where $S^{2}$ is variance, $\bar{\sigma}^{2}_{\rm err}$ is mean squared error and $\bar{X}^2$ is mean flux \citep{1990ApJ...359...86E,2003MNRAS.345.1271V}. The associated uncertainty in $F_{\rm var}$ is given by:

\begin{equation}
   \sigma_{F_{\mathrm{var}}} =  \sqrt{\left(\frac{1}{\sqrt{2N}}\frac{\bar{\sigma}^{2}_{err}}{F_{var}}\frac{1}{\bar{X}^2}\right)^2 + \left(\sqrt{\frac{\bar{\sigma}^{2}_{err}}{N}}\frac{1}{\bar{X}^2}\right)^2} 
\end{equation}
\citep{2015A&A...576A.126A,2018Galax...6....2B}.

\begin{figure}
\includegraphics[width=\columnwidth]{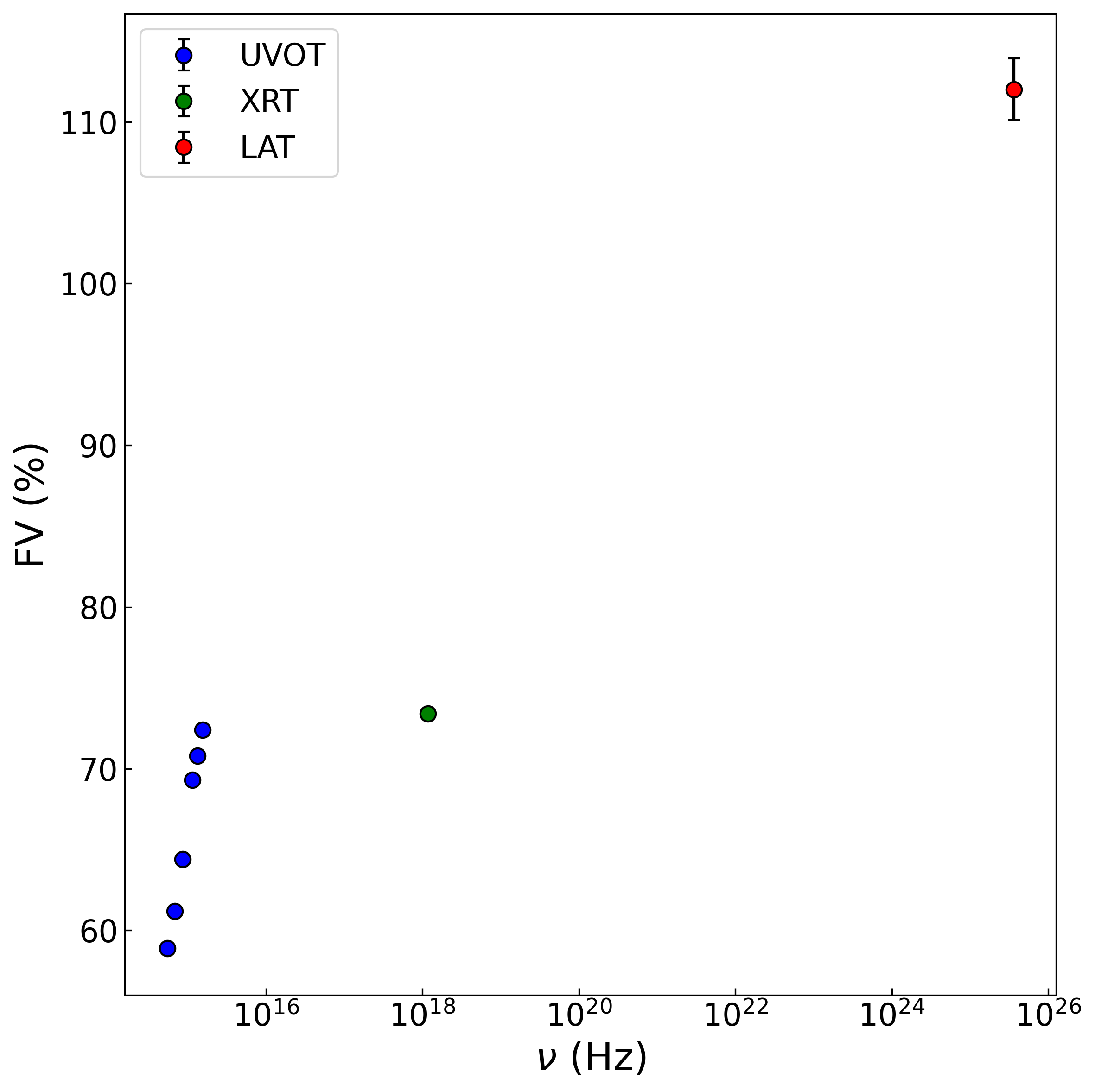}
\caption{Fractional variability (see Equation \ref{eq:fv}) of MWL lightcurves of the blazar OJ 287. 
\label{fig:fvfig}}
\end{figure}

The Figure \ref{fig:fvfig} and Table \ref{tab:fvtable} show the fractional variability of the multi-wavelength bands. The $\gamma$-rays show the highest variability, followed by X-rays and the UV and optical. 

\begin{table}
	\centering
	\caption{Fractional variability (in percent) of MWL lightcurves of the blazar OJ 287.}
	\label{tab:fvtable}
	\begin{tabular}{cc} 
		\hline
		Waveband & $F_{var} (\%)$\\
		\hline
        Swift UVOT: $v$ & 58.89 $\pm$ 0.15 \\
        Swift UVOT: $b$ & 61.22 $\pm$ 0.11 \\
        Swift UVOT: $u$ & 64.4 $\pm$ 0.11 \\
        Swift UVOT: $w1$ & 69.35 $\pm$ 0.11 \\
        Swift UVOT: $m2$ & 70.79 $\pm$ 0.13 \\
        Swift UVOT: $w2$ & 72.42 $\pm$ 0.11 \\
        Swift XRT & 73.41 $\pm$ 0.28 \\
        Fermi LAT & 111.77 $\pm$ 1.91 \\
		\hline
	\end{tabular}
\end{table}

\subsection{Correlation Analysis} \label{sec:corrzdcf}
\begin{figure*}
\centering
\includegraphics[width=0.3\textwidth]{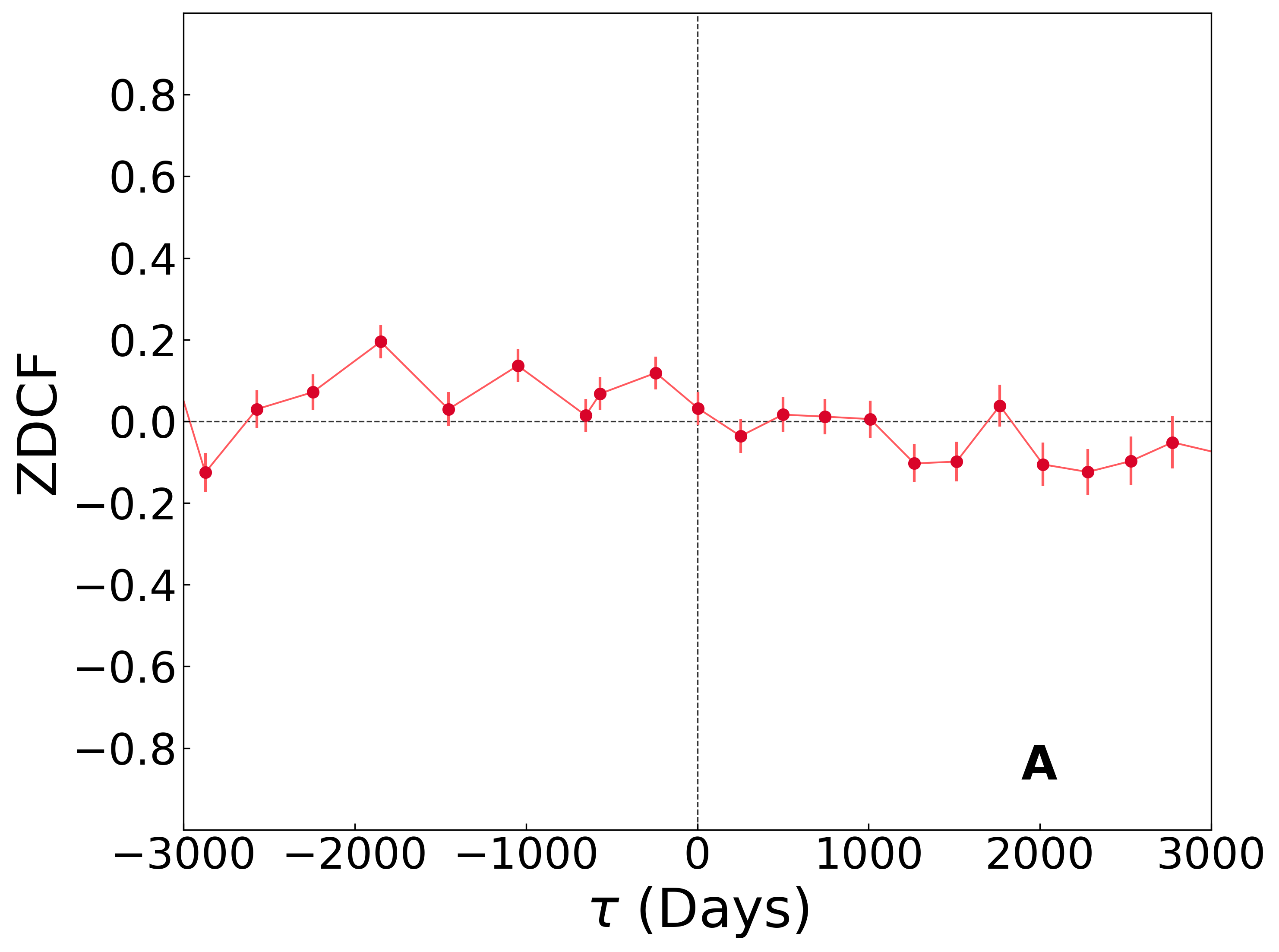}
\includegraphics[width=0.3\textwidth]{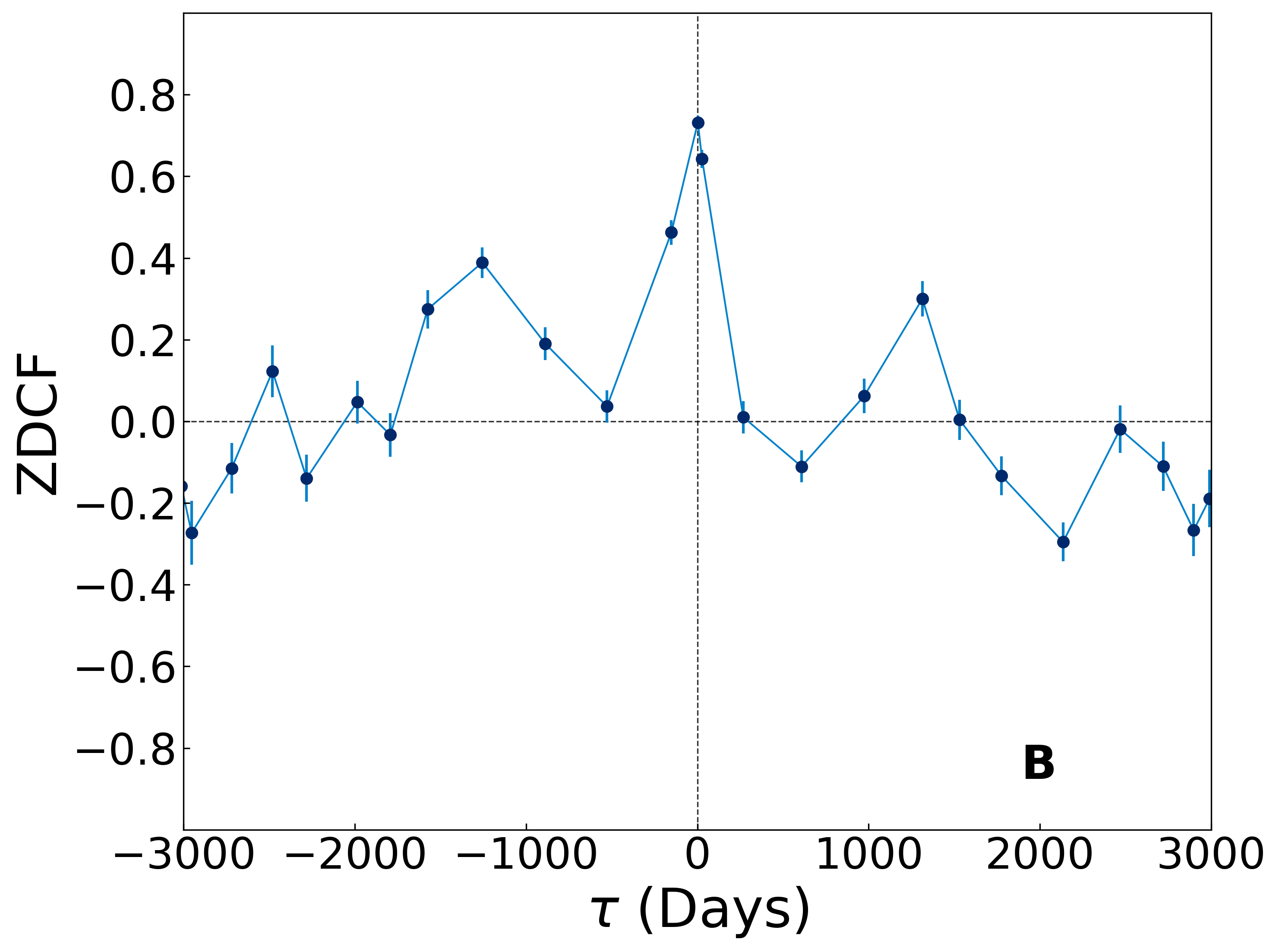}
\includegraphics[width=0.3\textwidth]{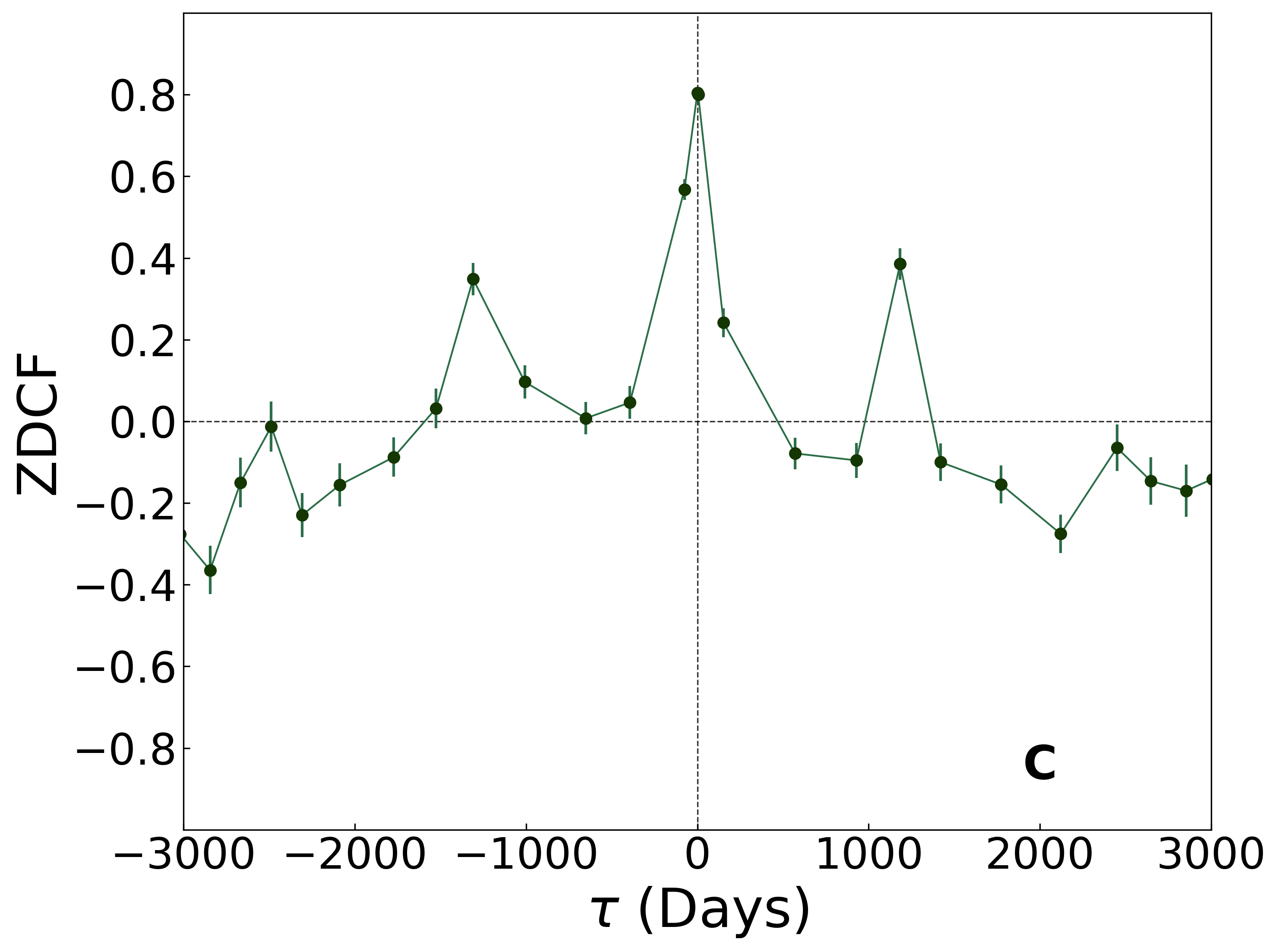}
\caption{The cross-correlation function estimated using ZDCF between MWL lightcurves in the above panels: (A) X-ray and $\gamma$-ray, (B) X-ray and Optical (V band), (C) X-ray and UV (W2 band).
\label{fig:zdcffig}}
\end{figure*}

Cross-correlation analysis between blazar lightcurves across various energy bands provides insights into the relationships between emission regions and underlying mechanisms in blazar jets, revealing temporal lags and spectral correlations that provide insights into jet physics. Due to the unevenly spaced time series data of the blazars, cross-correlation analysis is carried out using the method based on the discrete correlation function (DCF; \citealt{1988ApJ...333..646E}) over traditional correlation analysis. In this work, we implemented $z$-transformed Discrete Correlation Function (ZDCF; \citealt{1997ASSL..218..163A}). This algorithm implements Fisher's $z$-transformation to stabilize the skewed distribution of correlation coefficients, improving statistical performance when analyzing sparse data sets \citep{2021ApJ...923....7B}. The cross-correlation between X-ray and other energy bands, $\gamma$-ray, optical(V band) and UV (W2 band) was implemented by using pyZDCF \citep{jankov_isidora_2022_7253034}. The dense sampling of OJ 287 by Swift-XRT, UVOT, and Fermi-LAT made it possible to study and derive correlations among them. Figure \ref{fig:zdcffig} shows the cross-correlation between X-ray and other energy bands. We found a peak in the correlation between X-rays and the optical V band with a time lag of $2.82$ days and a correlation coefficient of $0.73$. For the X-ray and ultraviolet W2 band, the peak occurred at a time lag of -$0.05$ days with a correlation coefficient of $0.80$. The results reveal a significant correlation between the X-ray and optical/UV wavebands. However, no significant correlation was observed between the X-rays and $\gamma$-rays, consistent with the findings reported by \cite{2021A&A...654A..38P}.

\subsection{Epoch Selection} \label{sec:epoch}
\begin{figure*}
\includegraphics[width=\textwidth]{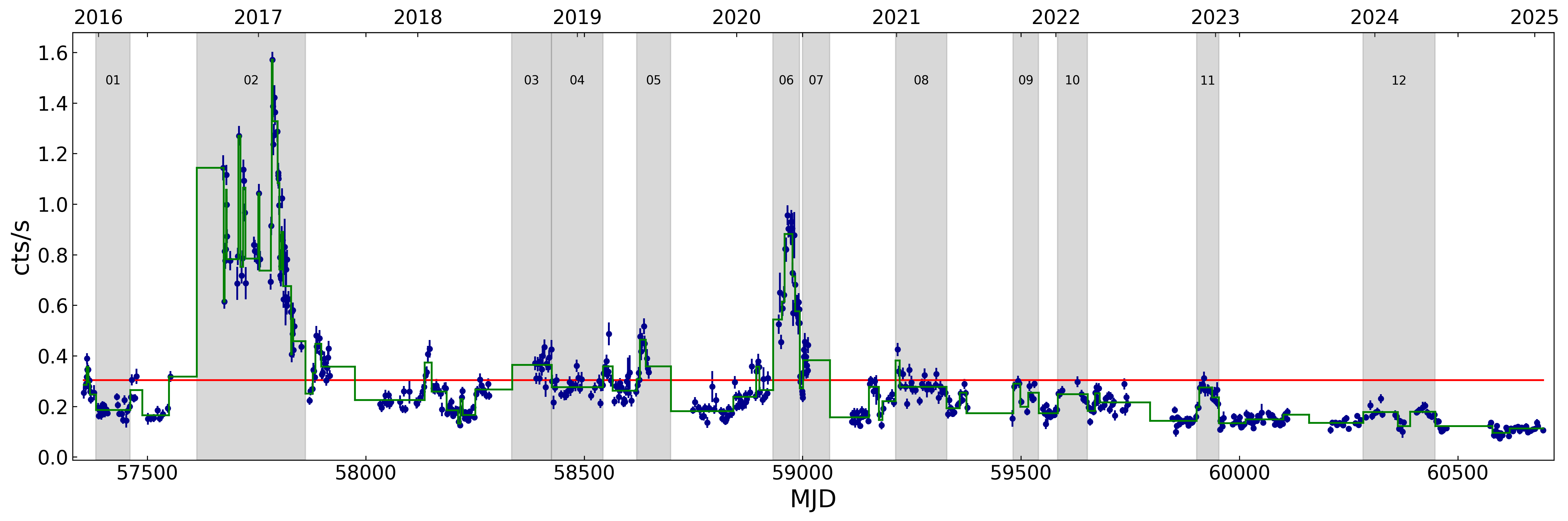}
\caption{X-ray lightcurve of OJ 287 from 2015 to 2016. The shaded regions represent different activity states used for SED modelling. The horizontal red line is the mean count rate ($0.304$ cts s$^{-1}$ with 1$\sigma$ uncertainty), while the green line shows the Bayesian Blocks analysis (see Section \ref{sec:epoch}).
\label{fig:xrtlc}}
\end{figure*}

The modelling of the MWL SEDs is a tool to probe blazars. It can provide detailed insight into different acceleration mechanisms. We have utilised data from Fermi-LAT, Swift-XRT and UVOT to produce MWL SED from optical/UV to $\gamma$-rays. A decade-long dataset, spanning from MJD 57382 to 60448, was used for the SED studies due to the availability of observations during different X-ray flux states: flaring, intermediate, and quiescent. For the epoch selection, we implemented Bayesian Blocks algorithm \citep{2013ApJ...764..167S}, which is a statistical method used for adaptive binning of data, particularly useful in time series analysis. This method allows for the identification of change points in data, enabling the detection of segments with different statistical properties. Bayesian Blocks provides a way to segment data into blocks based on a likelihood function that evaluates how well a proposed segmentation explains the observed data, where each block is characterized by a constant average value. We employed the Astropy\footnote{\url{https://docs.astropy.org/en/stable/api/astropy.stats.bayesian_blocks.html}} implementation of the Bayesian Blocks algorithm with a false-alarm probability of p$_0$ = 0.05 to identify different flux states in the X-ray lightcurve (see Figure \ref{fig:xrtlc}). To investigate the distribution and temporal evolution of the spectral parameters derived from the SED analysis, 12 representative epochs were carefully selected based on their X-ray flux characteristics and labeled Epochs 01–12 (see Table \ref{tab:epoch}).

\begin{table}
	\centering
	\caption{Details of the epochs estimated using Bayesian Blocks analysis (see Section \ref{sec:epoch}) selected for SED analysis.}
	\label{tab:epoch}
	\begin{tabular}{ccc} 
		\hline
		Epoch & Time Period & Duration (Days)\\
		\hline
		Epoch 01 & 57382-57460 & 78 \\
		Epoch 02 & 57613-57862 & 249 \\
		Epoch 03 & 58334-58425 & 91 \\
		Epoch 04 & 58425-58543 & 118 \\
		Epoch 05 & 58620-58698 & 78 \\
		Epoch 06 & 58932-58993 & 61 \\
		Epoch 07 & 59000-59062 & 62 \\
		Epoch 08 & 59213-59330 & 117 \\
		Epoch 09 & 59482-59540 & 58 \\
		Epoch 10 & 59584-59652 & 68 \\
		Epoch 11 & 59902-59953 & 51 \\
		Epoch 12 & 60283-60448 & 165 \\
		\hline
	\end{tabular}
\end{table}

\subsection{X-ray Spectral Variability}
\begin{figure}
\includegraphics[width=\columnwidth]{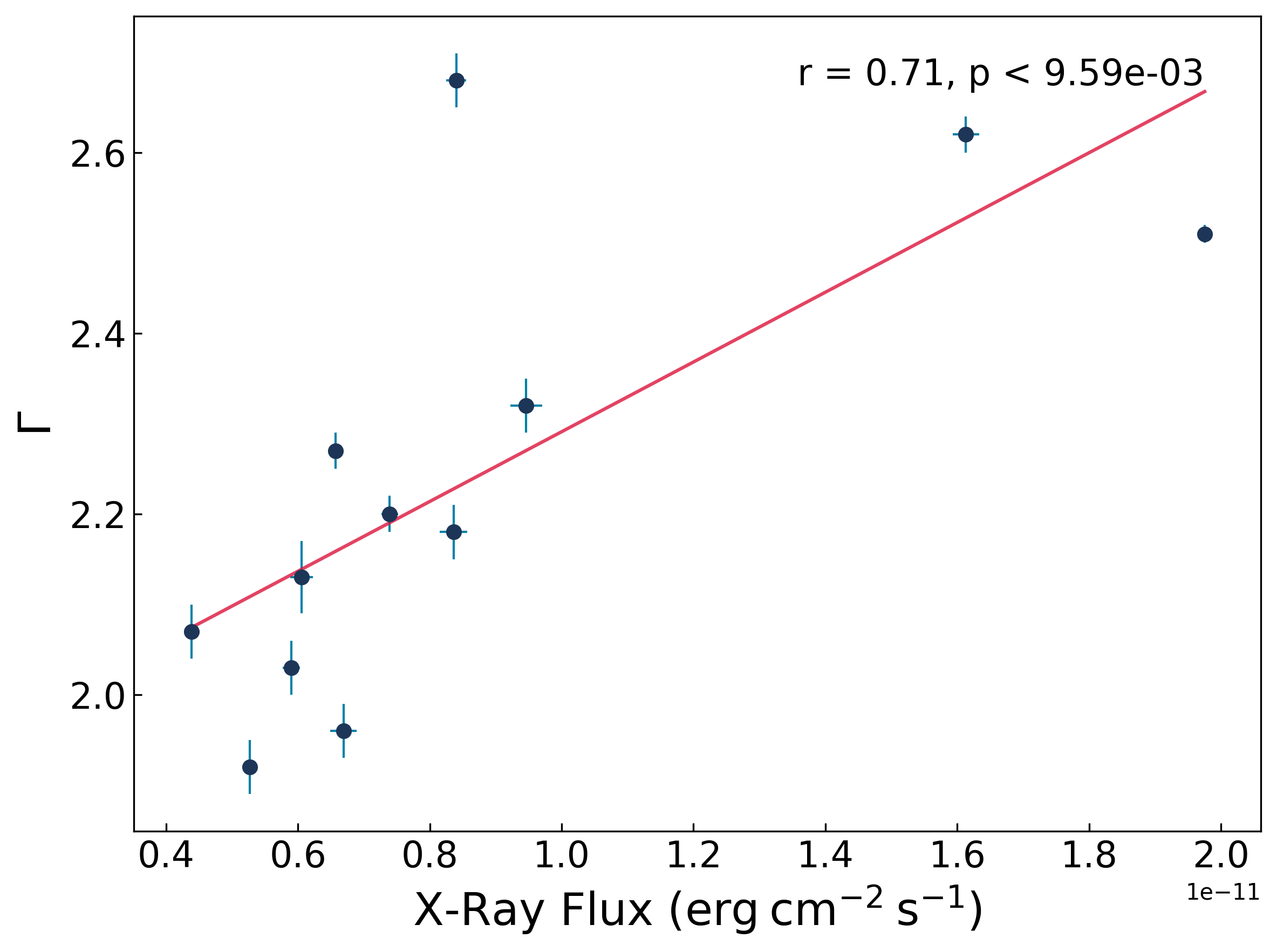}
\caption{Variation of photon index with X-ray flux for different activity states of the source OJ 287.
\label{fig:xraydist}}
\end{figure}

\begin{figure*}
\includegraphics[width=\textwidth]{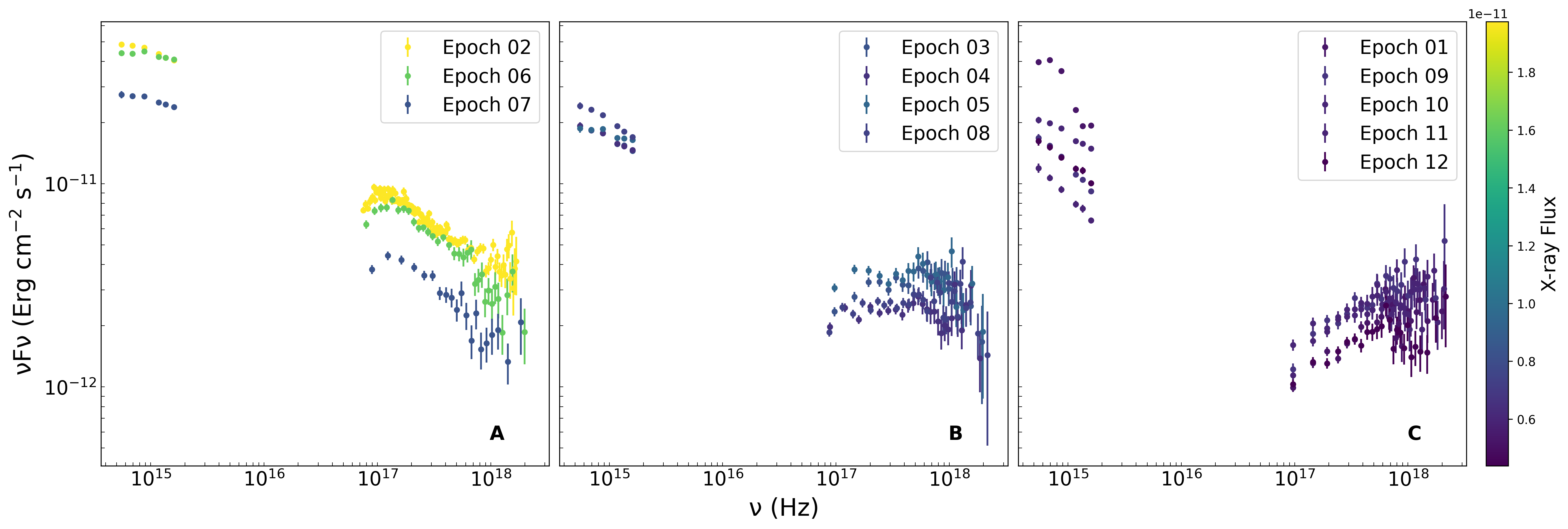}
\caption{Optical/UV to X-ray spectra during three distinct epochs characterized by X-ray flux states: (A) flare, (B) intermediate, and (C) quiescent.
\label{fig:xraysed}}
\end{figure*}

The source OJ 287 exhibits distinctive transitional behavior in its X-ray spectral properties, characterized by varying emission mechanisms that correlate with the source's flux states. During intermediate flux states, the X-ray emission comprises contributions from both synchrotron and inverse-Compton (IC) emission components, whereas during flaring events, the emission is predominantly attributed to the synchrotron component. Conversely, when the source transitions to quiescent states, the emission mechanism shifts primarily to the inverse-Compton process. This transitional nature of OJ 287 makes it a unique source among blazars, with only a few other sources exhibiting similar characteristics \citep{2021MNRAS.504.5575K, 2025MNRAS.536.1251W}. Table \ref{tab:xspec} presents the X-ray spectral analysis results obtained from XSPEC and Figure \ref{fig:xraydist} shows the photon index and X-ray flux, a clear positive correlation is seen with Pearson’s correlation coefficient $r = 0.71$ and ${\rm p-value} = 9.59\times10^{-3}$. The trimodal spectral behavior of the X-ray band is depicted in Figure \ref{fig:xraysed}. Panel (A) corresponds to high X-ray states (flaring), wherein the X-ray spectrum exhibits softness with photon indices $\Gamma > 2.51$. Panel (B) represents intermediate X-ray flux states, characterized by harder spectra with photon indices ranging from $2.18-2.32$. Panel (C) represents low X-ray flux states, displaying substantially harder spectra with photon indices $\Gamma < 2.1$. The spectral variations in the X-ray regime have been previously reported by \citep{2023AN....34420126K, 2020MNRAS.498L..35K}. This systematic transition from hard to soft X-ray spectra during flaring events is consistently observed throughout our SED analysis, confirming that the X-ray emission originates from the synchrotron component during flares and transitions to the inverse-Compton component as the source returns to quiescence. Figure \ref{fig:xraysed} further demonstrates the concurrent evolution of the optical/UV spectrum, which exhibits an inverse spectral hardening pattern relative to the X-ray behavior. The optical/UV spectrum appears softer when the source is in a quiescent state, progressively hardening as the source transitions through intermediate activity levels, and reaching maximum hardness during flaring episodes. The spectral variations in the optical/FUV regime have been previously studied by \citep{2022MNRAS.509.2696S}.

\begin{table}
	\centering
	\caption{X-ray spectral parameters from XSPEC analysis for the selected epochs.}
	\label{tab:xspec}
	\begin{tabular}{cccc} 
		\hline
		Epoch & Flux & $\Gamma$ & $\chi_{Red}^2$ \\
		   & $\rm{erg}\;cm^{-2} s^{-1}$ & & \\
		\hline
		Epoch 01 & 5.27e-12 $\pm$ 1.09e-13 & 1.92 $\pm$ 0.03 & 0.91 \\
		Epoch 02 & 1.98e-11 $\pm$ 7.50e-14 & 2.51 $\pm$ 0.01 & 1.19 \\
		Epoch 03 & 8.36e-12 $\pm$ 2.10e-13 & 2.18 $\pm$ 0.03 & 1.09 \\
		Epoch 04 & 6.57e-12 $\pm$ 1.03e-13 & 2.27 $\pm$ 0.02 & 0.94 \\
		Epoch 05 & 9.46e-12 $\pm$ 2.38e-13 & 2.32 $\pm$ 0.03 & 0.87 \\
		Epoch 06 & 1.61e-11 $\pm$ 2.00e-13 & 2.62 $\pm$ 0.02 & 0.84 \\
		Epoch 07 & 8.40e-12 $\pm$ 1.53e-13 & 2.68 $\pm$ 0.03 & 1.01 \\
		Epoch 08 & 7.39e-12 $\pm$ 1.25e-13 & 2.2 $\pm$ 0.02 & 0.95 \\
		Epoch 09 & 6.69e-12 $\pm$ 2.00e-13 & 1.96 $\pm$ 0.03 & 0.88 \\
		Epoch 10 & 6.06e-12 $\pm$ 1.71e-13 & 2.13 $\pm$ 0.04 & 0.85 \\
		Epoch 11 & 5.90e-12 $\pm$ 1.30e-13 & 2.03 $\pm$ 0.03 & 1.07 \\
		Epoch 12 & 4.38e-12 $\pm$ 9.45e-14 & 2.07 $\pm$ 0.03 & 0.99 \\
		\hline
	\end{tabular}
\end{table}

\subsection{Spectral Energy Distribution}
We implemented one-zone leptonic scenario for the modelling of broad band SEDs. The leptonic model assumes the relativistic leptons, mostly electrons and positrons interact with the magnetic field in emission region. They produce synchrotron photons which responsible for low-energy component of the SED from radio to X-rays.  The high-energy (HE) component of the SED is attributed to inverse Compton (IC) scattering of seed photons by this identical electron population. The IC process can happen through synchrotron-self Compton (SSC) or external Compton (EC) mechanisms based on the source of the seed photons. In SSC, the relativistic electrons up-scatter the same synchrotron photons that they produced in the magnetic field \citep{1996ApJ...461..657B, 1992ApJ...397L...5M, 1985A&A...146..204G}.

In the one-zone leptonic model, MWL emission is produced by a spherically symmetric blob of radius $R$, encompassing relativistic particles accelerated by various processes within an ambient magnetic field of strength $B$. The emission blob is oriented at an angle $\theta$ relative to the observer's line of sight and propagates along the jet with a bulk Lorentz factor $\Gamma$, which influences the emission region through the beaming factor $\delta = 1 / [\Gamma (1 - \beta \cos \theta)]$. This relativistic beaming amplifies the observed flux and shapes the spectral characteristics of the emitted radiation. Within the blob, relativistic electrons are distributed according to a broken power-law function
\begin{equation}
    N(\gamma) = N_0
    \begin{cases} 
        \gamma^{-p}, & \gamma_{min}\leq\gamma\leq\gamma_{br}, \\
        \gamma_{br}^{p_1-p} \gamma^{-p_1}, & \gamma_{br}\leq\gamma\leq\gamma_{max},
    \end{cases}
\end{equation}
where $\gamma_{min}$, $\gamma_{br}$, and $\gamma_{max}$ are the minimum Lorentz factor, broken Lorentz factor, and maximum Lorentz factor respectively. $p$ and $p_1$ are the spectral indices below and above the $\gamma_{br}$ and $N_0$ is the normalization constant in units of cm$^{-3}$.

The broadband SEDs of OJ 287 were modelled using JetSeT v1.2.2\footnote{\url{https://jetset.readthedocs.io/en/1.2.2/index.html}} tool. It is an open-source C/Python framework developed to fit numerical models to observed data for modelling emission mechanisms in relativistic astrophysical jets \citep{2020ascl.soft09001T, 2011ApJ...739...66T, 2009A&A...501..879T}. JetSeT implements numerical solutions to the particle transport equation to calculate emission spectra from various radiative processes.

MWL SED data are first imported and formatted into a JetSeT-compatible structure using the $ObsData$ module. Next, the parameter space is carefully constrained through the phenomenological modules $SEDShape$ and $ObsConstrain$, without performing a fit, while an electron distribution function is selected for the analysis. Subsequently, model optimization is performed using the $ModelMinimizer$ module. The phenomenological model is then fitted to the MWL SED data with available fitting methods, ensuring each free parameter remains constrained within physically meaningful boundaries. To improve convergence efficiency, we fix certain parameters such as redshift and the distance of the radiation region from the central black hole ($R_H$), while allowing the remaining parameters to vary within physically acceptable ranges. The quality of the fit is evaluated based on the $\chi^2/dof$ value.

\begin{figure*}
\centering
\subfigure{\includegraphics[width=0.3\textwidth]{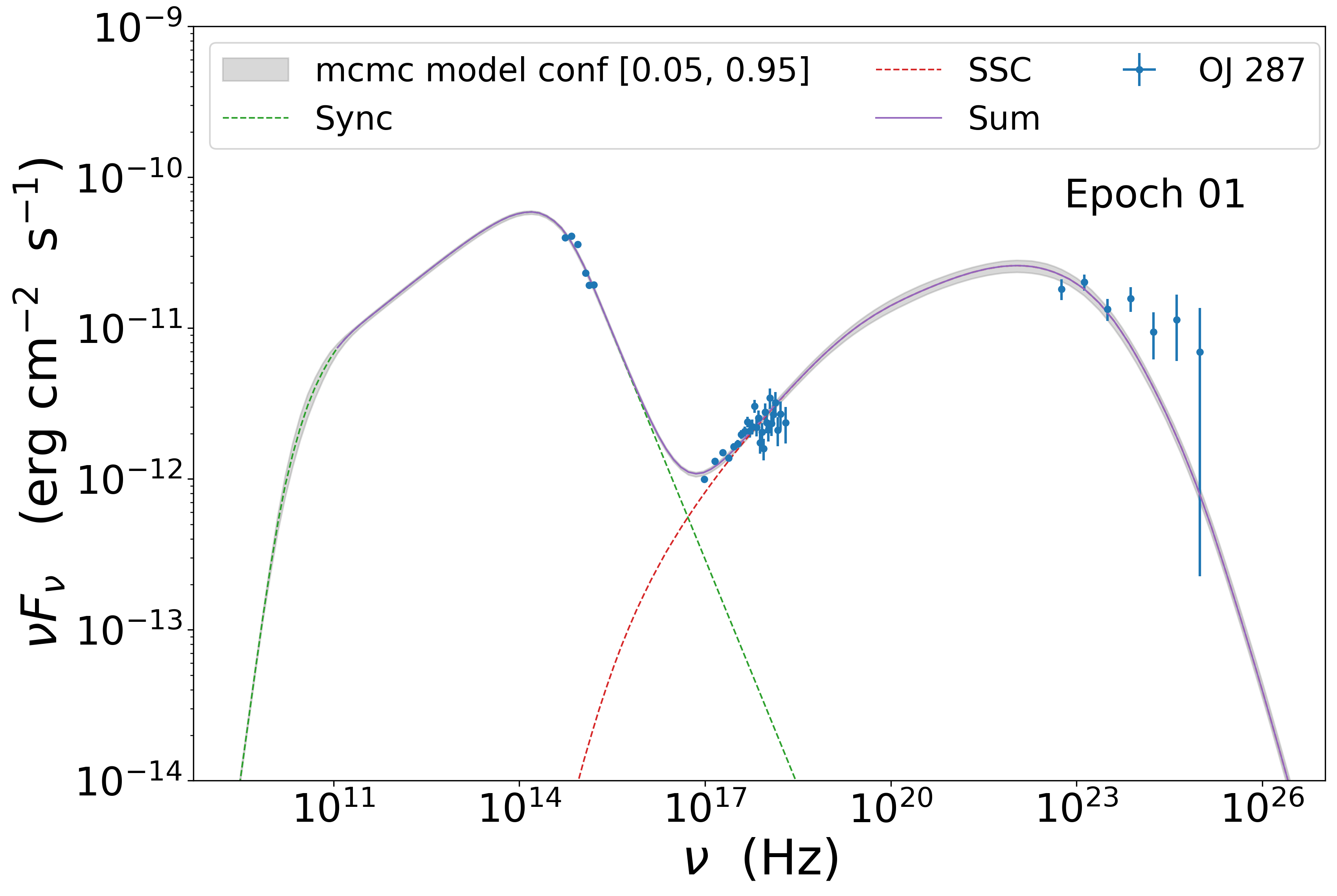}}
\subfigure{\includegraphics[width=0.3\textwidth]{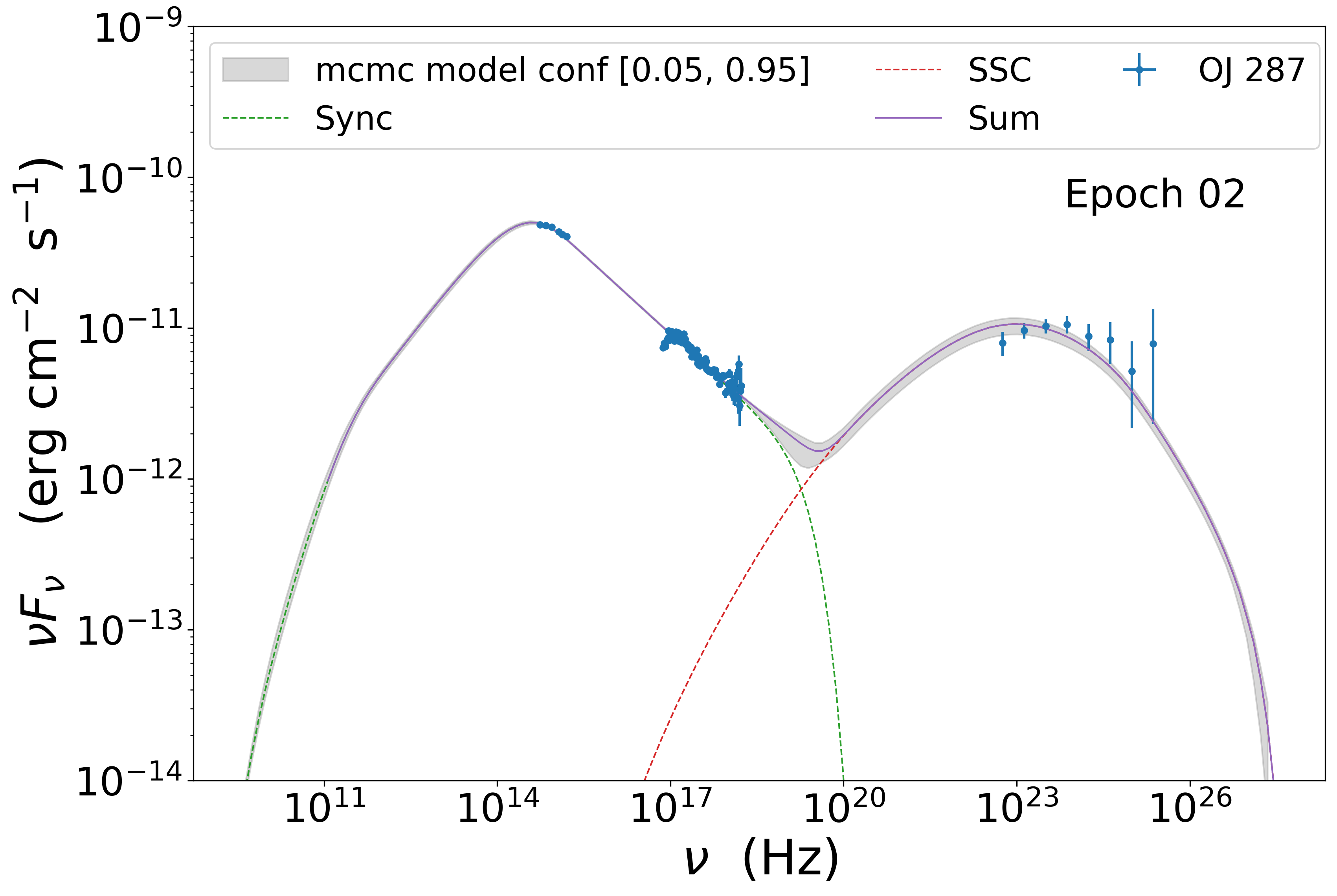}}
\subfigure{\includegraphics[width=0.3\textwidth]{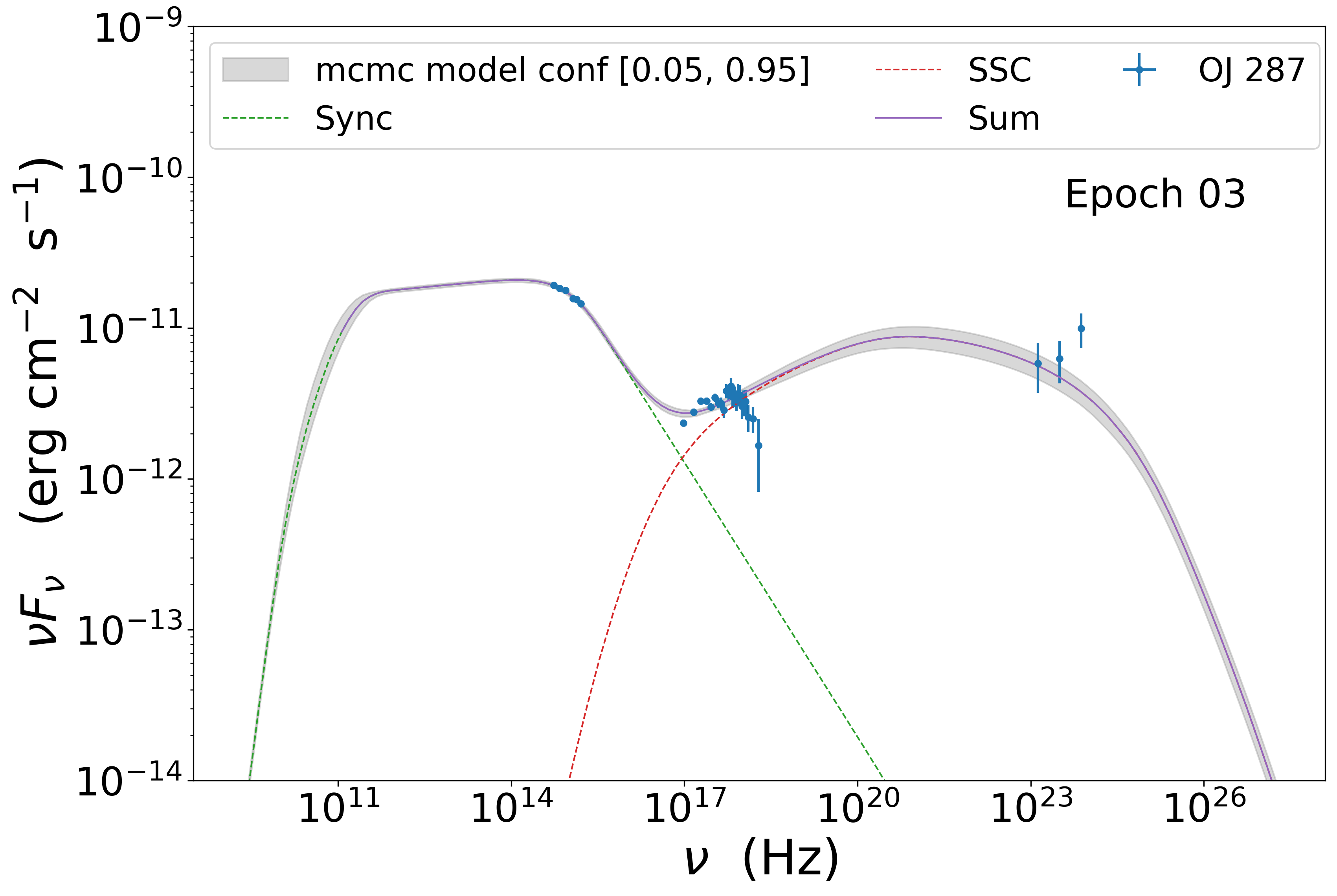}}
\subfigure{\includegraphics[width=0.3\textwidth]{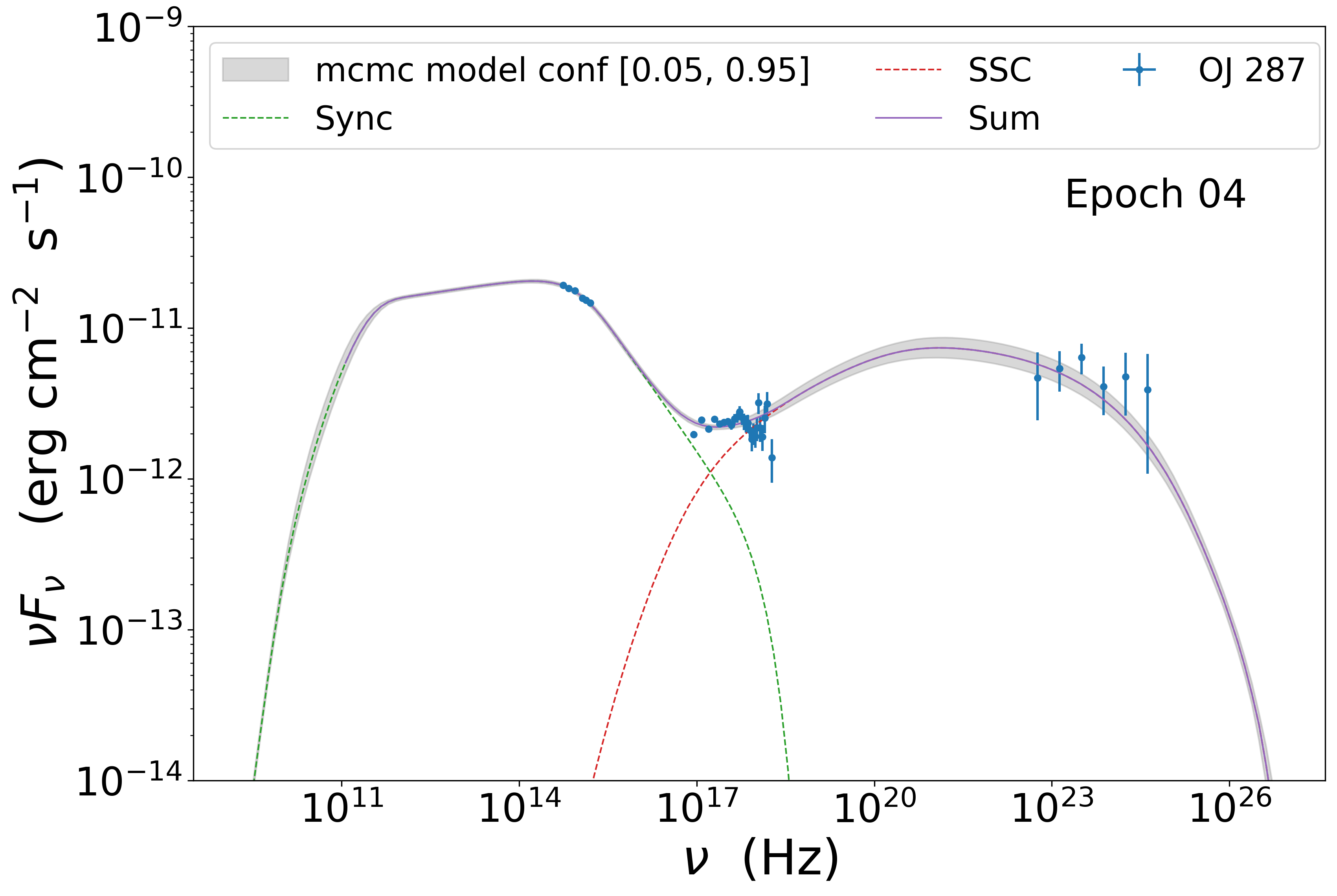}}
\subfigure{\includegraphics[width=0.3\textwidth]{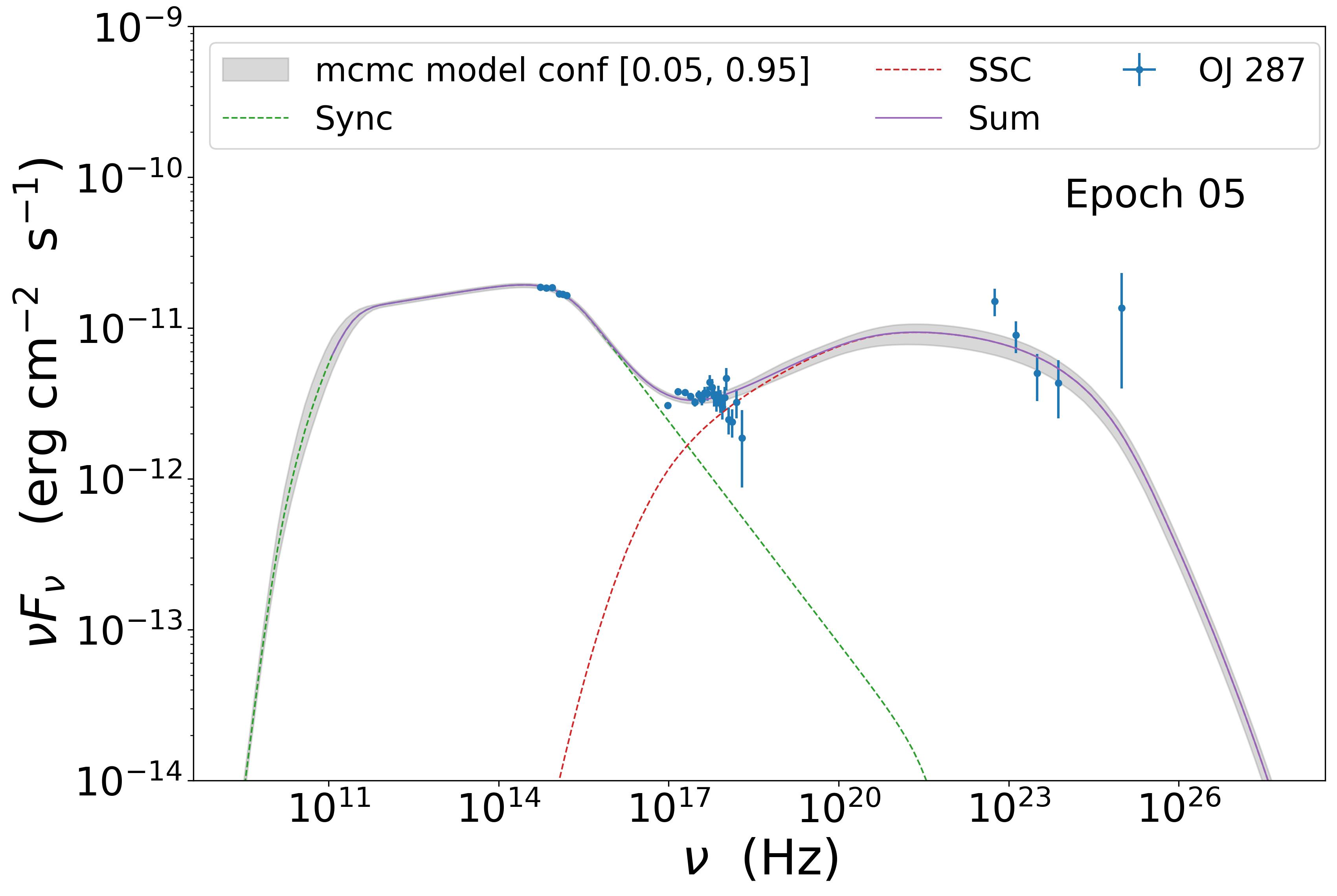}}
\subfigure{\includegraphics[width=0.3\textwidth]{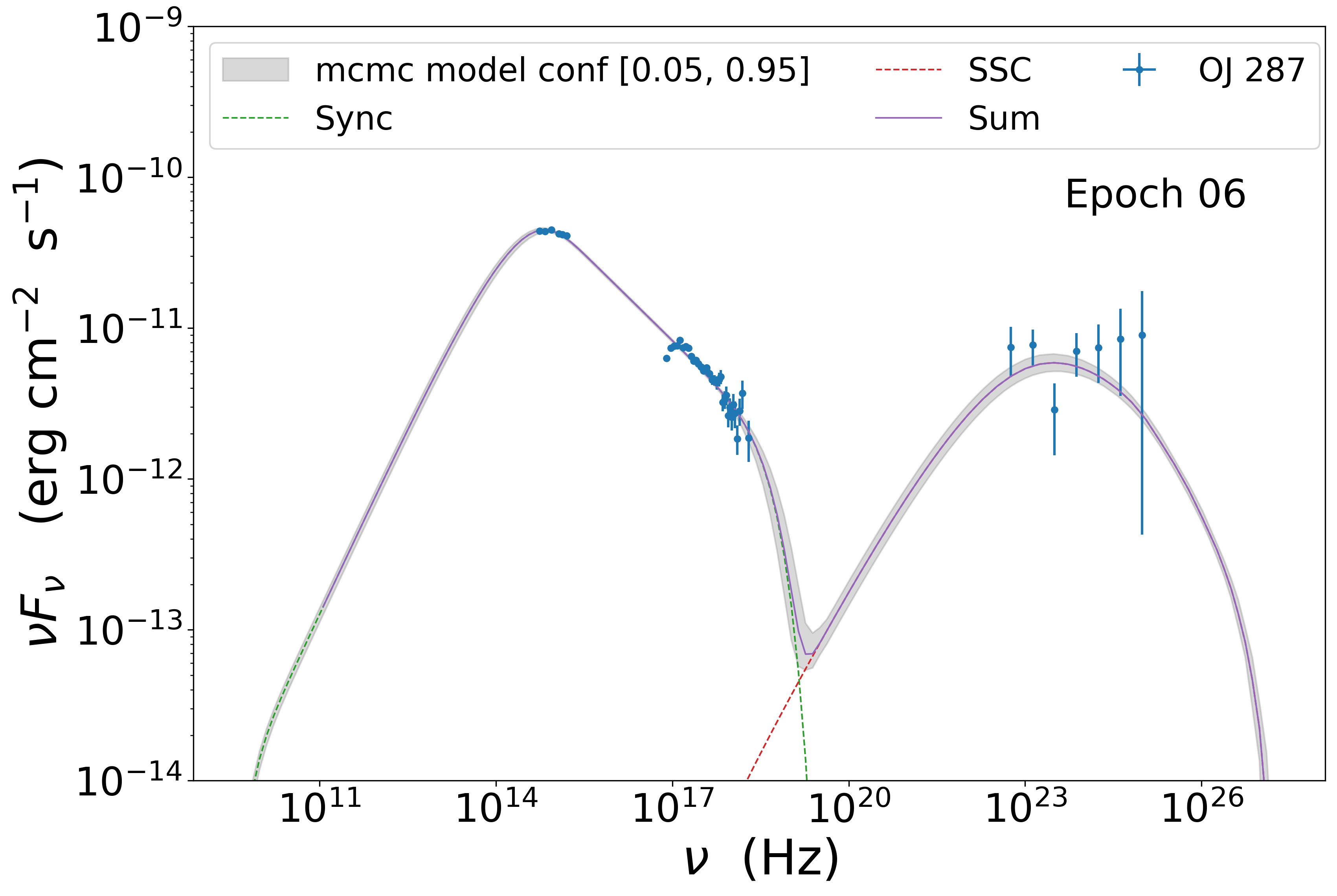}}
\subfigure{\includegraphics[width=0.3\textwidth]{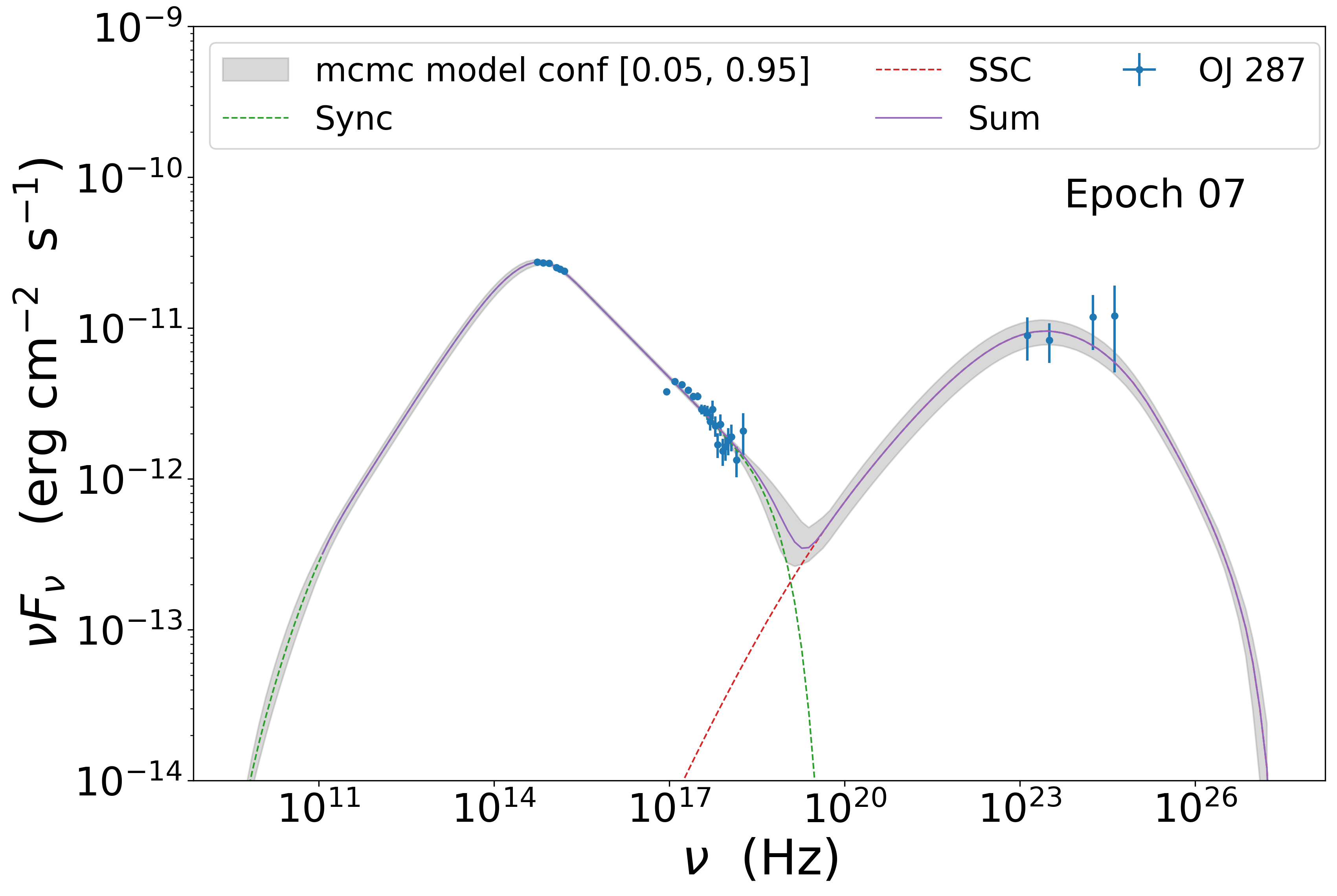}}
\subfigure{\includegraphics[width=0.3\textwidth]{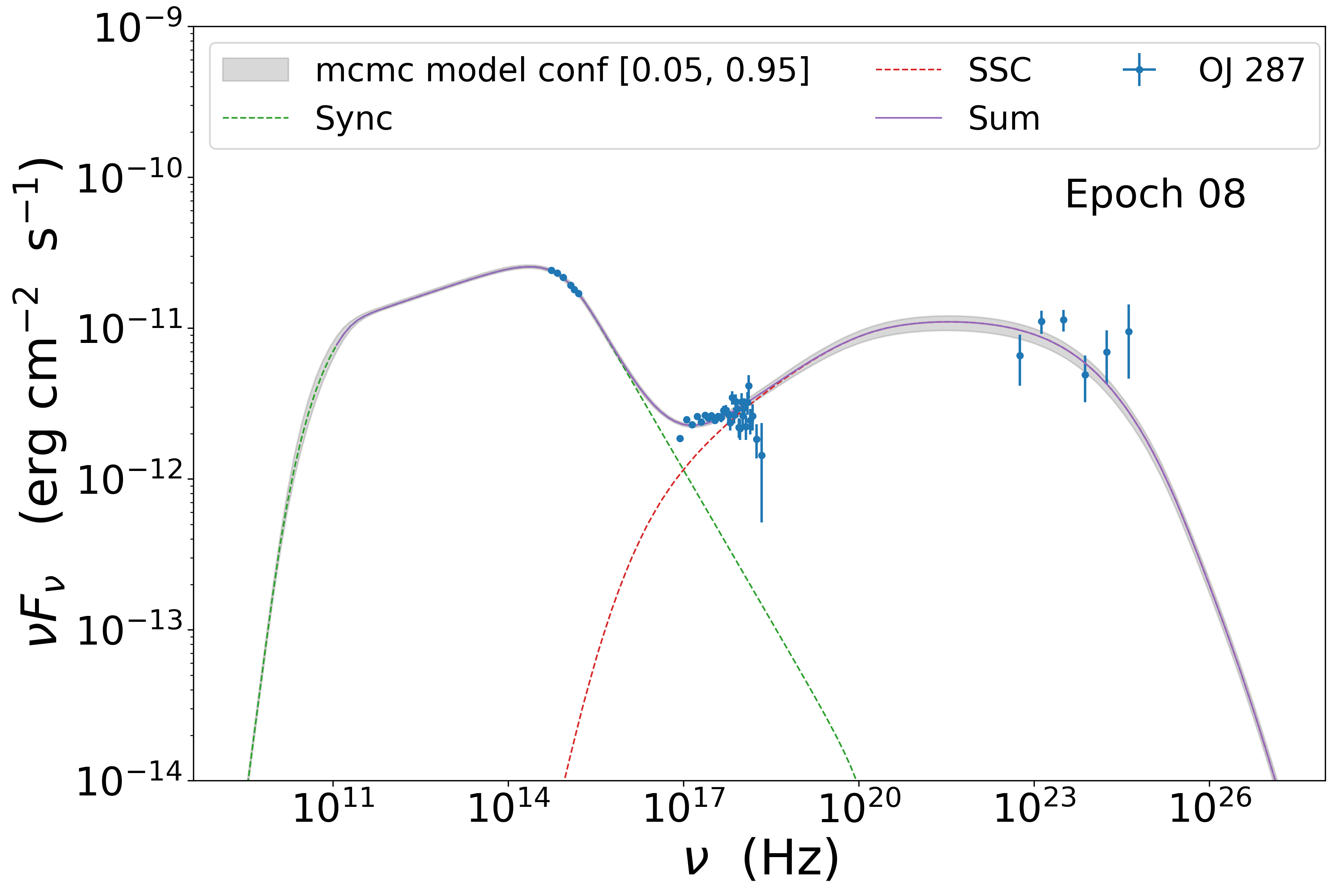}}
\subfigure{\includegraphics[width=0.3\textwidth]{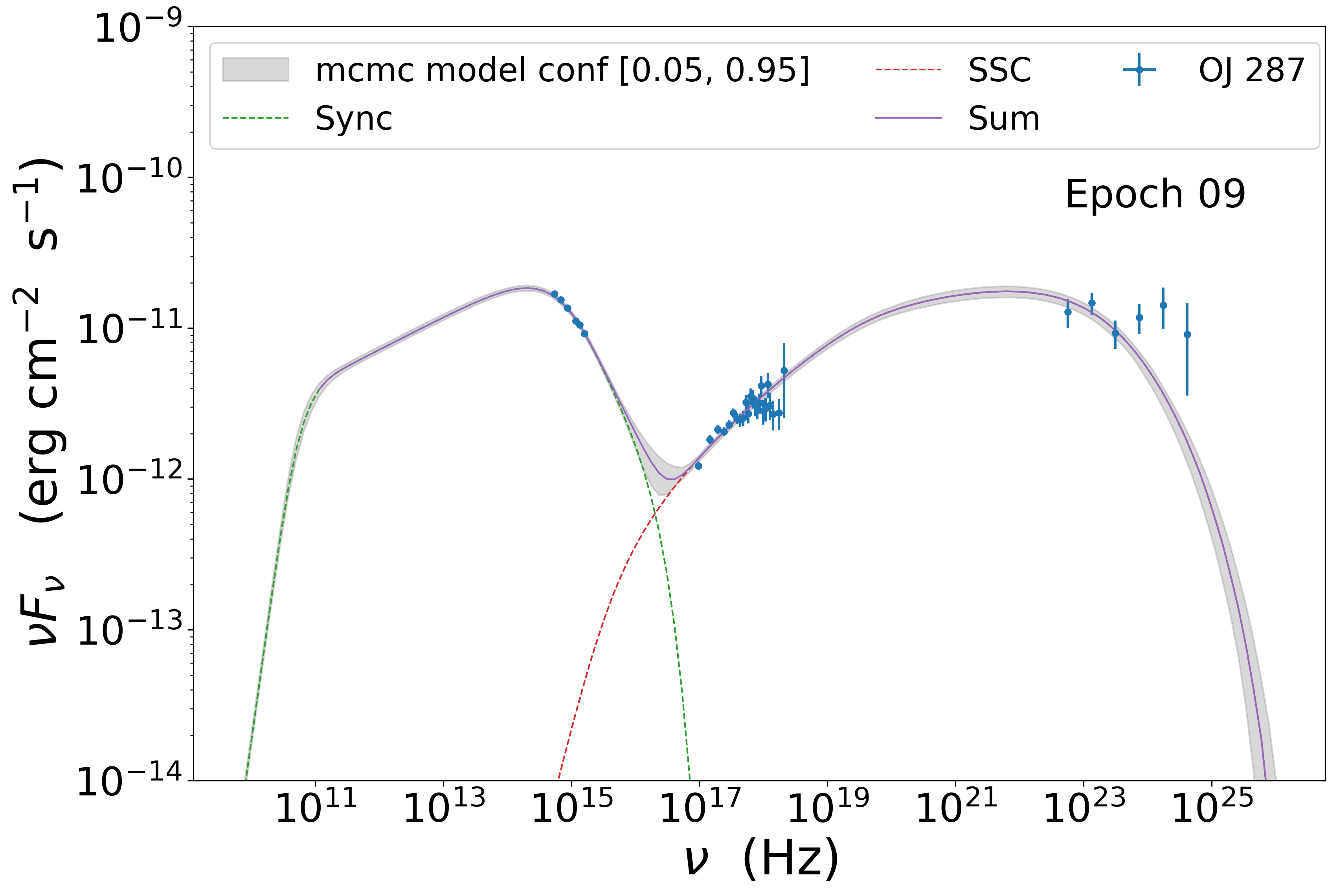}}
\subfigure{\includegraphics[width=0.3\textwidth]{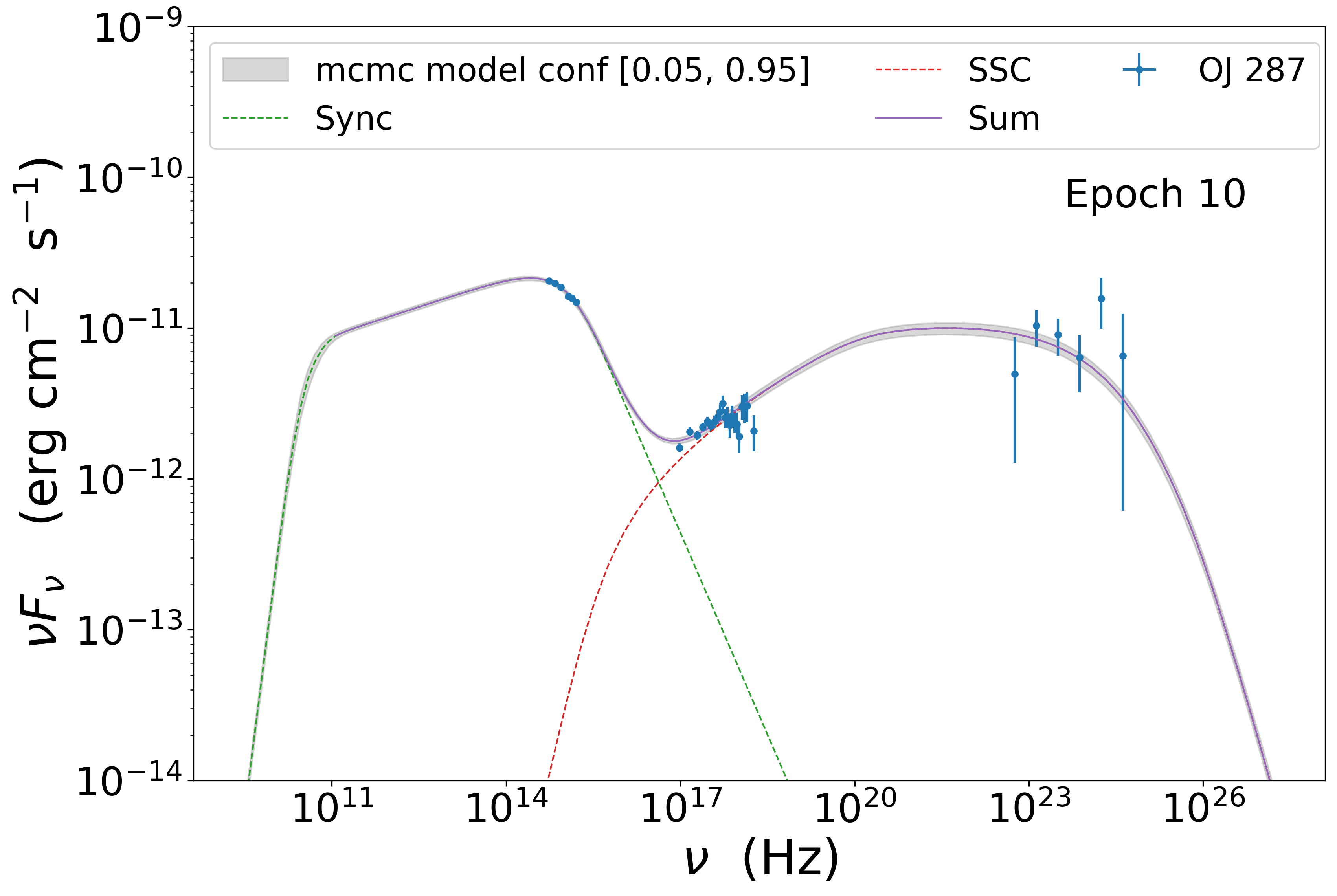}}
\subfigure{\includegraphics[width=0.3\textwidth]{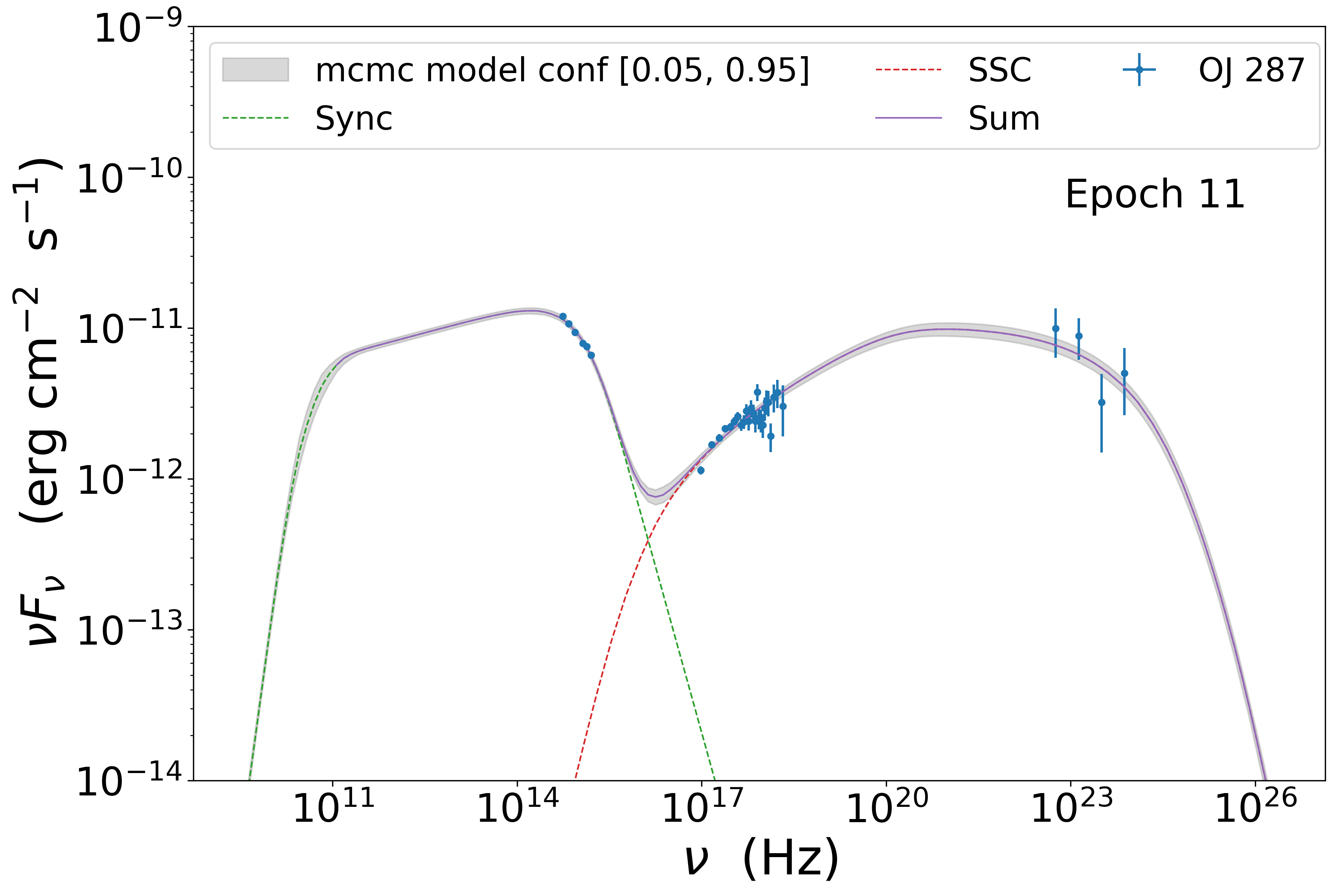}}
\subfigure{\includegraphics[width=0.3\textwidth]{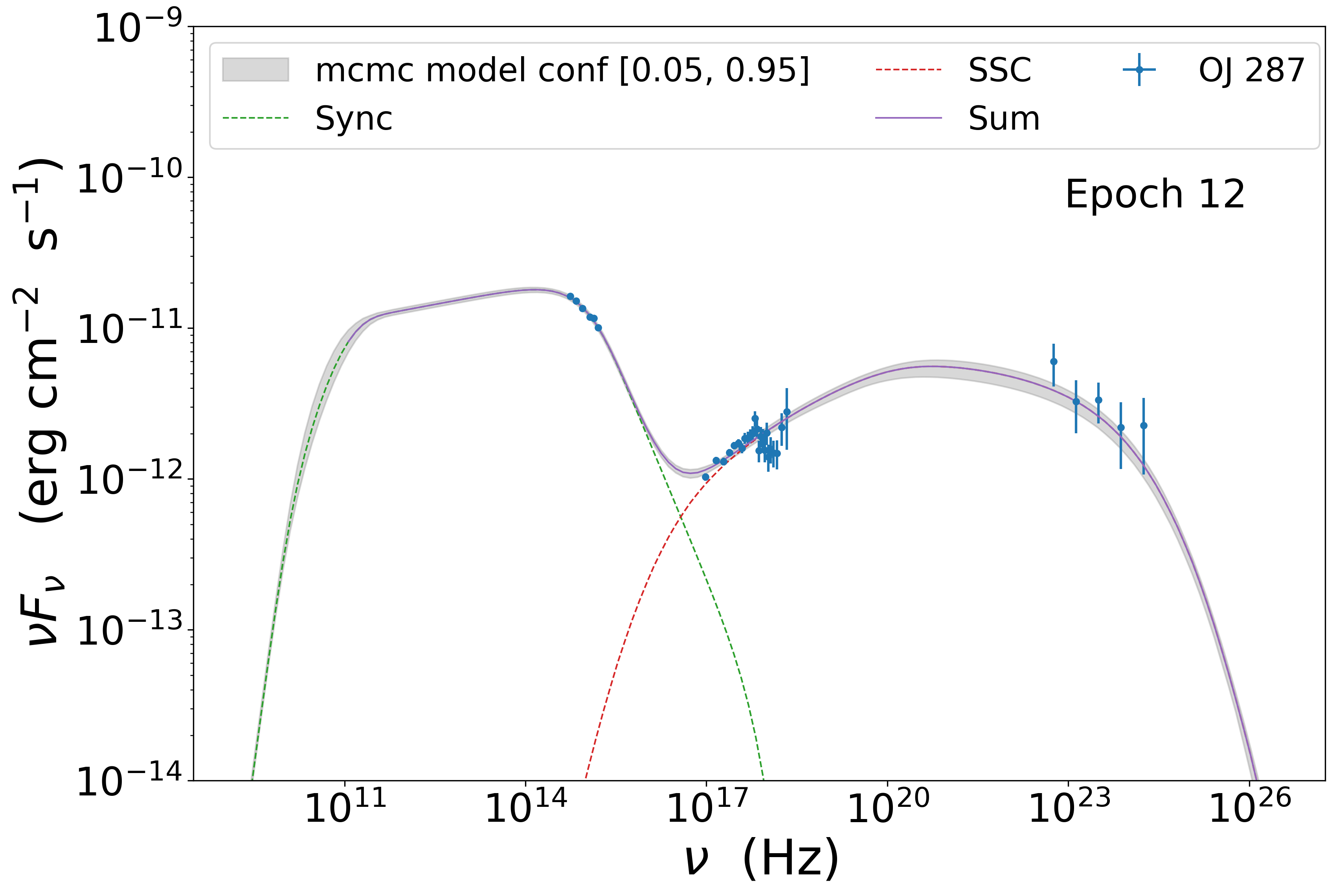}}
\caption{Broadband SEDs of OJ 287 modeled using a one-zone leptonic model. The solid line represents the best fit for the SEDs, while the green and red dashed lines represent emission from the synchrotron and SSC processes respectively, with the grey shaded region indicating the uncertainty region derived from MCMC sampling of the model parameters.
\label{fig:sed}}
\end{figure*}

\begin{table*}
	\centering
	\caption{Best-fit model parameters for the broadband SEDs of OJ 287.}
	\label{tab:sedfit}
	\begin{tabular}{ccccccccccc} 
		\hline
		Epoch & $\gamma_{min}$ & $\gamma_{max}$ & $N$ & $\gamma_{br}$ & $p$ & $p_1$ & $R$ & $B$ & $\delta$ & $\chi_{Red}^2$ \\
		\hline
		Epoch 01 & 197.84 & 6.50e+07 & 1.37 & 1.51e+04 & 2.35 & 5.0 & 4.79e+17 & 1.59e-02 & 24.64 & 4.95 \\
		Epoch 02 & 254.41 & 2.37e+06 & 0.35 & 1.04e+04 & 2.0 & 3.7 & 2.56e+17 & 4.30e-02 & 24.14 & 3.7 \\
		Epoch 03 & 308.81 & 4.22e+07 & 1.54 & 2.71e+04 & 2.92 & 4.21 & 4.80e+17 & 1.59e-02 & 23.41 & 2.8 \\
		Epoch 04 & 325.92 & 7.25e+05 & 1.4 & 2.19e+04 & 2.88 & 4.12 & 3.99e+17 & 2.44e-02 & 20.93 & 2.45 \\
		Epoch 05 & 295.86 & 5.42e+07 & 1.81 & 2.79e+04 & 2.88 & 3.98 & 4.61e+17 & 2.84e-02 & 16.66 & 2.54 \\
		Epoch 06 & 1.06 & 9.88e+05 & 0.33 & 1.19e+04 & 1.33 & 3.76 & 2.57e+17 & 4.32e-02 & 25.25 & 4.22 \\
		Epoch 07 & 180.15 & 1.61e+06 & 0.37 & 1.44e+04 & 1.77 & 3.79 & 2.17e+17 & 3.59e-02 & 21.27 & 3.07 \\
		Epoch 08 & 247.83 & 1.03e+07 & 2.43 & 2.25e+04 & 2.74 & 4.33 & 3.67e+17 & 1.66e-02 & 25.07 & 3.42 \\
		Epoch 09 & 152.95 & 8.36e+04 & 29.69 & 1.58e+04 & 2.56 & 4.71 & 1.17e+17 & 2.09e-02 & 28.07 & 1.98 \\
		Epoch 10 & 203.77 & 7.95e+07 & 5.54 & 3.68e+04 & 2.75 & 4.79 & 2.30e+17 & 2.86e-03 & 71.43 & 1.98 \\
		Epoch 11 & 270.24 & 4.80e+08 & 8.51 & 3.42e+04 & 2.78 & 5.93 & 1.70e+17 & 3.51e-03 & 59.28 & 1.91 \\
		Epoch 12 & 262.85 & 5.88e+05 & 1.4 & 2.53e+04 & 2.85 & 4.92 & 4.93e+17 & 1.56e-02 & 23.3 & 2.09 \\
		\hline
	\end{tabular}
	{\footnotesize
	\parbox{\linewidth}{\textbf{Notes --}
	$\gamma_{\min}$: Minimum Lorentz factor; 
	$\gamma_{\max}$: Maximum Lorentz factor; 
	$N$: Emitters density (cm$^{-3}$); 
	$\gamma_{\rm br}$: Break Lorentz factor; 
	$p$: Low‐energy spectral index; 
	$p_{1}$: High‐energy spectral index; 
	$R$: Region size (cm); 
	$B$: Magnetic field (G); 
	$\delta$: Doppler factor.}}
\end{table*}

Figure \ref{fig:sed} shows the best fit broadband SED of all the epochs and Table \ref{tab:sedfit} shows the best fit parameters. The SED plots show that OJ 287 exhibits three distinct spectral behaviors in its X-ray emission. During quiescent states, the X-ray emission is contributed by SSC processes. In intermediate activity states, the X-ray emission displays contributions from both SSC and synchrotron emission mechanisms. When the source is in its flaring activity, the X-ray spectrum is characterized exclusively by synchrotron emission. These observations clearly show a consistent transitional spectral behavior of OJ 287 in the X-ray regime.

\begin{figure*}
\includegraphics[width=\textwidth]{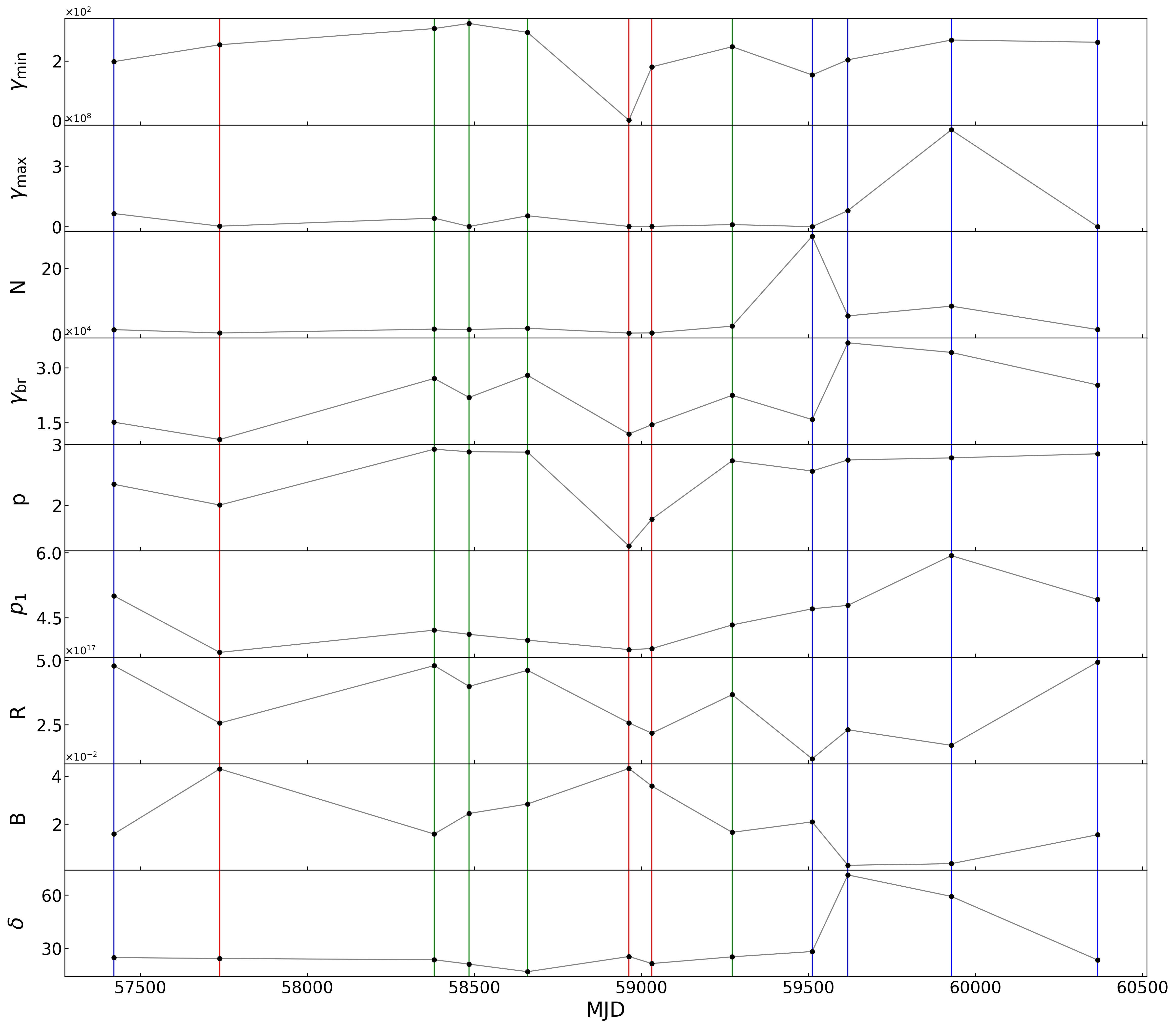}
\caption{Jet parameter variations derived from SED modeling across different X-ray activity epochs. Vertical lines are the midpoints of epochs with red representing flare states, blue for quiescent states, and green for intermediate states.
\label{fig:paramvar}}
\end{figure*}

Figure \ref{fig:paramvar} shows the temporal evolution of the jet parameters across all epochs. The vertical lines in the figure represent different epochs, color-coded according to their X-ray activity states: synchrotron (flaring), transition (intermediate), and SSC (quiescent) \citep{2025arXiv250418927M}. The $\gamma_{min}$ exhibits relatively consistent behavior across epochs, with notably lower values during flaring states compared to intermediate and quiescent states. In contrast, the $\gamma_{max}$ shows distinct patterns: quiescent states display significant variability, while flaring and intermediate states maintain comparatively stable, lower values. The $N$ follows a similar trend, with variations exclusively in quiescent states. The $\gamma_{br}$ demonstrates a clear trend across activity states: flaring epochs consistently show lower values, intermediate states exhibit moderate values, and quiescent states generally display higher values. Both spectral slopes show state-dependent behavior: $p$ remains below 2.0 during flaring epochs, while increasing to $2.73-2.92$ in intermediate states and $2.34-2.84$ in quiescent states. Similarly, the $p_1$ is lower during flaring states ($<3.8$) compared to intermediate ($3.98-4.33$) and quiescent states ($4.7-5.93$). The $R$ exhibits clear correlation with activity states: the flare states have smaller region sizes ($2.16-2.57 \times 10^{17}$ cm), intermediate states consistently maintain larger region sizes ($3.67-4.80 \times 10^{17}$ cm), while quiescent states show significant variations ($1.17-4.93 \times 10^{17}$ cm). The $B$ increases dramatically during flaring states ($3.6 \times 10^{-2} - 4.3 \times 10^{-2}$ G), decreases in intermediate states ($1.6 \times 10^{-2} - 2.8 \times 10^{-2}$ G), and reaches its lowest values during quiescent states ($3.0 \times 10^{-3} - 2.1 \times 10^{-2}$ G), demonstrating a strong inverse correlation with the emission region size. Finally, the  $\delta$ remains relatively stable across flaring and intermediate states, but shows significant variability in quiescent states.

\begin{figure}
\includegraphics[width=\columnwidth]{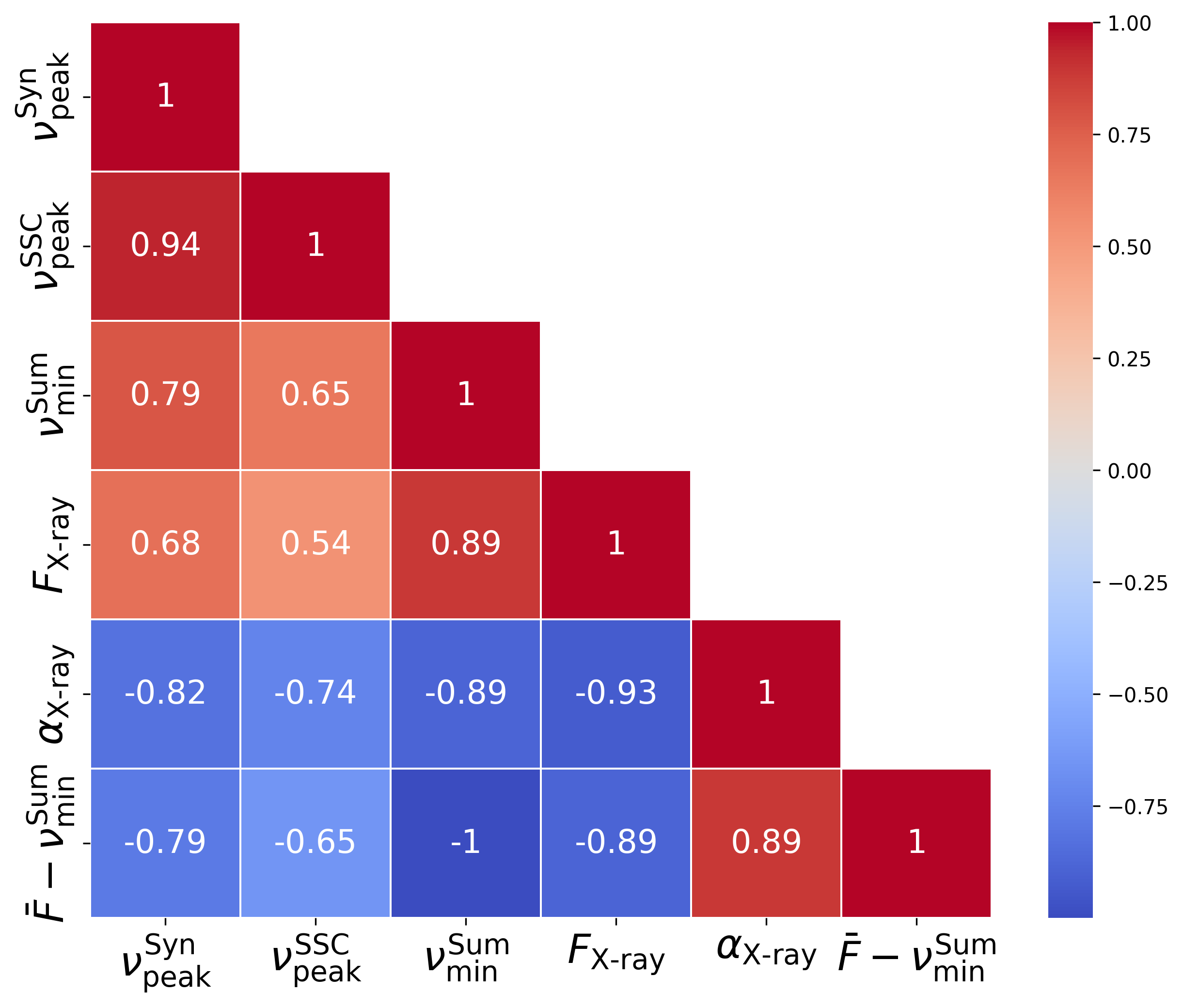}
\caption{Correlation matrix showing the relationships between parameters derived from SED modeling.
\label{fig:corrmatrix}}
\end{figure}

From the SED model and X-ray spectrum, we have calculated the synchrotron peak frequency ($\nu_{\text{peak}}^{\text{syn}}$), SSC peak frequency ($\nu_{\text{peak}}^{\text{SSC}}$), transition frequency between synchrotron and SSC components ($\nu_{\text{min}}^{\text{Sum}}$), X-ray flux ($F_{\text{X-ray}}$), X-ray spectral slope ($\alpha_{\text{X-ray}}$), and the frequency difference between X-ray mean flux and transition point ($F_{\text{X-ray}} - \nu_{\text{min}}^{\text{Sum}}$, the inflection point between synchrotron and IC peaks). To analyse the relation between these parameters, a correlation matrix was computed using the Pandas\footnote{\url{https://pandas.pydata.org/docs/reference/api/pandas.DataFrame.corr.html}} module and shown in Figure \ref{fig:corrmatrix}. This matrix contains the pairwise Pearson correlation coefficients between each parameter in our dataset, measuring the strength and direction of their linear relationships. The matrix shows significant relationships between the parameters: The $\nu_{\text{peak}}^{\text{Syn}}$ shows a very strong positive correlation ($r = 0.941$) with the ($\nu_{\text{peak}}^{\text{SSC}}$, indicating these two peaks tend to shift together. This supports that a tight connection between the two emission components in the SED. Similarly, $\nu_{\rm peak}^{\rm Syn}$ also correlates positively with $\nu_{\rm min}^{\rm Sum}$ ($r = 0.79$) and with $F_{\rm X-ray}$ ($r = 0.68$), indicating that higher synchrotron peak frequencies are associated with both higher transition frequencies and increased X-ray flux. The $\nu_{\text{min}}^{\text{Sum}}$ demonstrates a strong positive correlation with X-ray flux ($r = 0.886$), indicating that with higher transition frequencies tend to be brighter in X-rays. $F_{\text{X-ray}} - \nu_{\text{min}}^{\text{Sum}}$ shows a negative correlation with $F_{\text{X-ray}}$ ($r = -0.89$) indicating that when the source is active in X-rays, $F_{\text{X-ray}} - \nu_{\text{min}}^{\text{Sum}}$ is large and transition point tend to move to higher energies. Negative correlation ($r = -0.99$) is found between $\nu_{\rm min}^{\rm Sum}$ and $\bar{F} - \nu_{\rm min}^{\rm Sum}$, which is expected due to the mathematical dependence between these variables. The positive correlation ($ r = 0.89$) between $\alpha_{\rm X-ray}$ and $\bar{F} - \nu_{\rm min}^{\rm Sum}$ supports the idea that spectral shape in the X-ray band is linked with the positioning of the transition point in the SED.

\section{Discussion} \label{sec:discussion}
OJ 287 exhibits strong flux variability across all observed wavelength regimes, a characteristic well-established in previous studies highlighting its dynamic nature \citep{2021Univ....7..261K, 2023arXiv230516144K}. We found strong correlation patterns among the optical, ultraviolet (UV), and X-ray bands. The optical and UV bands are consistently reported to be closely correlated, often with near-zero lag times \citep{2021Univ....7..261K, 2021ApJ...923...51K, 2018MNRAS.479.1672K, 2018MNRAS.473.1145K, 2020ApJ...890...47P, 2021A&A...654A..38P}, indicating that the emission in these bands likely originates from the same or closely linked physical processes and regions, possibly the low-energy end of the synchrotron spectrum \citep{2023arXiv230516144K}. The X-ray and optical/UV emission are more complex and appears to depend on the source's activity state. Our finding of strong correlation patterns among these bands aligns with periods when the X-ray and optical/UV variations are simultaneous or closely lagged \citep{2018MNRAS.479.1672K}. However, previous studies have also reported epochs where the correlation is less clear or where distinct lag/lead patterns emerge \citep{2017MNRAS.468..426S, 2018MNRAS.479.1672K, 2020ApJ...890...47P}. For instance, during quiescence periods, X-rays have been found to lead or lag the optical-UV, which has been interpreted as evidence for a different X-ray emission component dominating during low states, likely Inverse Compton (IC) emission. Conversely, during outbursts, X-rays tend to follow the UV with near-zero lags, suggesting a stronger link between the emission processes during high-activity periods \citep{2021ApJ...923...51K}.

The observation of two prominent X-ray flares in 2017 and 2020 with clear optical/UV counterparts is consistent with previous detailed studies of these events \citep{2021Univ....7..261K, 2021ApJ...923...51K, 2022MNRAS.513.3165K, 2022MNRAS.515.2778H, 2021ApJ...921...18K}. These outbursts were characterised by exceptionally high flux levels and distinct spectral properties, particularly a strong soft X-ray excess and a 'softer-when-brighter' behavior in the X-ray band \citep{2021ApJ...923...51K, 2021MNRAS.504.5575K, 2021MNRAS.508..315P}. This spectral behavior implies that a synchrotron component becomes increasingly dominant during high states. The detection of these flares across optical, UV, and X-ray bands supports a common or closely related origin for the emission in these regimes during outburst states, potentially involving particle acceleration and cooling processes within the jet. The 2016-2017 activity, which included the first VHE detection of OJ 287,  was notably associated with a new, additional HBL-like emission component seen in the optical-to-X-ray spectrum, which revived during the 2020 outburst\citep{2018MNRAS.479.1672K, 2018MNRAS.479.1672K, 2022JApA...43...79K, 2022MNRAS.509.2696S}. This changing contribution of different spectral components: synchrotron, SSC, potentially external Compton or even thermal components during specific states, \citep{2021MNRAS.504.5575K, 2023arXiv230516144K, 2020Galax...8...15K} likely underlies the variations in correlation strength and lag patterns observed between optical/UV and X-rays across different activity states. The gamma-ray emission shows no statistically significant correlation with other wavebands throughout our 16-year monitoring period. This is highlighted by the detection of optical/UV and X-ray flares in 2015, 2017 and 2020 without detectable signatures in gamma-ray band, and conversely, a prominent gamma-ray outburst in 2011 (MJD 55809.0 to 55900.0) that lacked counterparts in other wavebands. This lack of overall correlation, despite prominent flaring in multiple bands, suggests that the primary emission regions or mechanisms driving the variability in gamma-rays may be decoupled from those in the optical-to-X-ray bands over long timescales \citep{2018MNRAS.479.1672K}. 

The broadband SED of OJ 287 has been extensively modeled across various observational periods using different theoretical frameworks. Our study focuses on investigating the transitional nature of the X-ray spectrum of OJ 287 and one-zone leptonic model is implemented. While the broadband SED of OJ 287 can exhibit complex spectral behavior that necessitates multi-zone modeling approaches \citep{2018MNRAS.479.1672K, 2024ApJ...973..134A}, a one-zone leptonic model can adequately reproduce the observed data when the analysis is focused on X-ray spectral transitions \citep{2021A&A...654A..38P}. Previous studies have employed different modelling depending on the specific observational phenomena under investigation. \cite{2013MNRAS.433.2380K} implemented a one-zone model incorporating synchrotron, SSC, and EC processes to analyze the 2009 flare. \cite{2018MNRAS.479.1672K} implemented a two-zone framework, with distinct emission regions responsible for low-synchrotron-peaked (LSP) and high-synchrotron-peaked (HSP) spectral characteristics. The 2015 December flare was modelled by \cite{2018MNRAS.473.1145K} using a one-zone approach that included synchrotron, SSC, and EC scattering processes. An alternative hadronic interpretation was proposed by \cite{2020MNRAS.498.5424R} for the 2015 November flare, wherein high-energy emission was attributed to proton-proton interactions within the binary black hole impact outflow scenario. \cite{2021A&A...654A..38P} employed a single-zone SSC model utilizing time-dependent MWL data spanning 2017-2020, motivated by the observed correlated variability across all wavebands indicating a co-spatial emission origin. \cite{2024ApJ...973..134A} adopted a multi-zone model to account for very high-energy (VHE) $\gamma$-ray observations, while \cite{2025MNRAS.tmp..668H} introduced a neural network-based SSC model to systematically analyze time-resolved SEDs.

The SED analysis reveals a distinctive and consistent transitional behavior in the X-ray spectral properties of OJ 287, which correlates directly with the source's flux states. Specifically, during quiescent periods, the X-ray emission is primarily attributed to IC processes, transitioning to a mix of both synchrotron and IC components during intermediate flux states, and becoming predominantly synchrotron-driven during flaring events. This systematic transition from harder to softer X-ray spectra as the source enters flaring states, and returns to harder spectra in quiescence, is a key characteristic observed in our SED analysis. This finding is consistent with and builds upon previous MWL studies of OJ 287, which have long noted its highly variable nature across different wavebands including significant X-ray spectral variability \citep{2021Univ....7..261K, 2021MNRAS.504.5575K, 2021ApJ...923...51K, 2021A&A...654A..38P, 2023arXiv230516144K}. The X-ray spectrum of OJ 287 has been observed to span a wide range of states, from very flat to exceptionally steep \citep{2021MNRAS.504.5575K, 2021A&A...654A..38P, 2020ApJ...890...47P}. Our results strongly support the "softer-when-brighter" pattern observed in the X-ray emission of OJ 287 \citep{2021MNRAS.504.5575K}. The observation that the X-ray spectrum becomes softer during flaring episodes, reaching photon indices $\Gamma > 2.51$ in high states, points to the increased dominance of the synchrotron component at these times. The optical/UV spectrum also exhibits correlated behavior, showing an inverse hardening pattern relative to the X-rays, appearing softer in quiescence and hardening during flares. This simultaneous spectral behavior across the optical/UV and X-ray bands underscores the connection between the synchrotron emission in these regimes \citep{2018MNRAS.479.1672K, 2023arXiv230516144K}.

The observed transitional behavior of OJ 287's X-ray emission across different activity states can be physically interpreted through changes in the relative dominance of the synchrotron and inverse Compton (IC) emission components within the blazar's relativistic jet. The changes in the relative contributions of the synchrotron and IC components are driven by variations in the physical conditions within the blazar jet \citep{2021A&A...654A..38P}. The observed X-ray spectral transition in OJ 287, particularly the shift towards synchrotron dominance during flares, is best explained by flaring activity within the relativistic jet that increases the contribution of the synchrotron emission in the X-ray band \citep{2021MNRAS.504.5575K, 2023AN....34420126K}. The temporal evolution of the model parameters, as captured by modelling the SEDs over time, is crucial for understanding the physical processes driving the observed variability. The high amplitude of parameter changes over time directly reflects dynamic changes in the physical conditions within the emitting region of the blazar jet \citep{2025MNRAS.tmp..668H}. These parameter variations, such as changes in the electron energy distribution, magnetic field strength, or Doppler factor dictate the relative contributions of synchrotron and IC emission, leading to the observed spectral transitions.

The systematic variation of these parameters across different activity states provides compelling evidence for distinct physical conditions governing each spectral state of OJ 287 \citep{2021A&A...654A..38P, 2025MNRAS.tmp..668H}. During flaring states, the combination of stronger magnetic fields, lower broken Lorentz factors, and softer spectral indices creates ideal conditions for synchrotron emission to dominate the X-ray band. The particles contributing to flaring states may be accelerated to high energies through various acceleration mechanisms, such as shock acceleration, stochastic acceleration, and magnetic reconnection, within the turbulent blazar jet \citep{1987PhR...154....1B, 2009MNRAS.395L..29G, 1998A&A...333..452K, 1998A&A...335..134O}. Conversely, during low activity states, weaker magnetic fields coupled with harder spectral slopes favour inverse Compton processes, resulting in SSC-dominated X-ray emission. The intermediate states, characterized by moderate values of most parameters, naturally produce the observed transitional spectrum with contributions from both emission mechanisms. The correlation analysis of the SED parameters provides quantitative support for this physical picture. The strong positive correlation ($r = 0.941$) between the synchrotron peak frequency ($\nu_{\text{peak}}^{\text{Syn}}$) and the SSC peak frequency ($\nu_{\text{peak}}^{\text{SSC}}$) is expected in leptonic models where both components originate from the same electron population within a single or closely related emission region.  The positive correlations found between $\nu_{\rm peak}^{\rm Syn}$, the transition frequency between components ($\nu_{\rm min}^{\rm Sum}$), and the X-ray flux ($F_{\rm X-ray}$) are particularly insightful. As the source becomes brighter in X-rays (flaring state), the synchrotron peak frequency tends to increase, and the transition frequency between the components shifts towards higher energies. This supports the interpretation that during flares, the synchrotron emission component becomes more prominent and its peak shifts to higher energies, including or dominating the X-ray band \citep[see also][]{2025MNRAS.536.1251W,2025JHEAp..4700365N}. 

\section{Summary} \label{sec:summary}
This paper presents a comprehensive MWL study of the blazar OJ 287 from 2008 September to 2025 January, utilising data from the Swift (optical/UV, X-ray) and Fermi ($\gamma$-ray) observatories. Our analysis focused on characterising the variability, correlations, and spectral evolution across X-ray band and modelling the SEDs. The main results of the work can be summarised as follows: 

\begin{itemize}
    \item Long-Term MWL Variability and Flaring: OJ 287 exhibits significant flux variations in optical, UV, X-ray, and $\gamma$-ray bands throughout the monitoring period, demonstrating strong MWL variability. Two major X-ray flares in 2017 and 2020 were observed alongside optical/UV counterparts but lacked associated $\gamma$-ray activity. In contrast, a major $\gamma$-ray flare occurred without counterparts in other bands.

    \item X-ray Spectral Transition: The X-ray spectrum showed a clear flux-dependent transition. In low-flux (quiescent) states, it was hard ($\Gamma < 2.1$), consistent with inverse Compton emission. During intermediate flux levels ($\Gamma = 2.18$--$2.32$), mixed synchrotron and IC components were present. High-flux (flaring) states exhibited soft spectra ($\Gamma > 2.51$), dominated by synchrotron emission, confirming the ``softer-when-brighter" behavior.
    
    \item Spectral Modelling and Correlation Analysis: One-zone leptonic SED modelling captured the spectral evolution, supporting the role of changing dominance between synchrotron and IC processes. Correlation matrices revealed strong links between synchrotron/IC peaks and X-ray flux and photon index, reinforcing the interpretation of the spectral transitions.
\end{itemize}
In conclusion, this MWL analysis of OJ 287 reveals the complexity of blazar emission mechanisms, including differential flaring behavior and flux-dependent spectral evolution. Applying a similar approach to a collective study of a sample of blazars would further shed light on the dynamic nature of particle acceleration and radiation processes within relativistic jets.

\section*{ACKNOWLEDGMENTS}
This work was partially supported by a program of the
Polish Ministry of Science under the title “Regional Excellence Initiative,” project No. RID/SP/0050/ 2024/1. We sincerely appreciate the anonymous reviewer for their thoughtful review.

\section*{Data Availability}
The $\gamma$-ray data used in this paper are publicly available through the Fermi Science Support Center (FSSC) at NASA's Goddard Space Flight Center. The X-ray and optical/UV data are also publicly available from the archives of HEASARC, maintained by NASA.

\bibliographystyle{mnras}
\bibliography{mnras} 


\bsp	
\label{lastpage}
\end{document}